\documentclass[graybox]{svmult}

\usepackage{mathptmx}       
\usepackage{helvet}         
\usepackage{courier}        
\usepackage{type1cm}        
\usepackage{makeidx}         
\usepackage{graphicx}        
\usepackage{multicol}        
\usepackage[bottom]{footmisc}

\usepackage{fullpage}

\makeindex             

\begin{document}

\title*{Formation of the First Galaxies: Theory and Simulations}
\author{Jarrett L. Johnson}
\institute{Jarrett L. Johnson \at Max-Planck-Institut f{\"u}r extraterrestrische Physik, 
Giessenbachstra\ss{}e, 85748 Garching, Germany \email{jjohnson@mpe.mpg.de} \\
Los Alamos National Laboratory, Los Alamos, NM 87545, USA}
%
%
\maketitle


\abstract{The properties of the first galaxies are shaped in large part by the first generations of stars, 
which emit high energy radiation and unleash both large amounts of mechanical energy and the first heavy elements when they explode as supernovae.  
We survey the theory of the formation of the first galaxies in this context, focusing on the results of cosmological simulations to 
illustrate a number of the key processes that define their properties.  We first discuss the evolution of the primordial gas as it is incorporated into 
the earliest galaxies under the influence of the high energy radiation emitted by the earliest stars; we then turn to consider how the injection of 
heavy elements by the first supernovae transforms the evolution of the primordial gas and alters the character of the first galaxies.
Finally, we discuss the prospects for the detection of the first galaxies by future observational missions, in particular focusing on
the possibility that primordial star-forming galaxies may be uncovered.}

\section{Introduction: Defining Characteristics of the First Galaxies}
While the first stars are for the most part well-defined objects, the definition of the first galaxies is somewhat more ambiguous (see e.g. Bromm \& Yoshida 2011).  Here we shall 
adopt the common view that a galaxy must be able to host ongoing star formation, even in the face of the radiative and mechanical feedback that accompanies the formation and evolution of stars.   
By this definition, the formation sites of the first stars, dark matter minihalos with masses 10$^5$ - 10$^6$ M$_{\odot}$, are unlikely candidates for the first galaxies, as the high
energy radiation emitted by young stars and the supernovae that mark their end of life can rarify and expel any dense gas from which stars may form at a later time.  As shown in Figure 1, it is only
somewhat larger halos, with masses 10$^7$ - 10$^8$ M$_{\odot}$, which have deep enough gravitational potential wells and enough mass to prevent the expulsion of gas after an episode of star formation 
(e.g. Kitayama \& Yoshida 2005; Read et al. 2006; Whalen et al. 2008).  

\begin{figure}[t]
\sidecaption[t]
\includegraphics[scale=1.3]{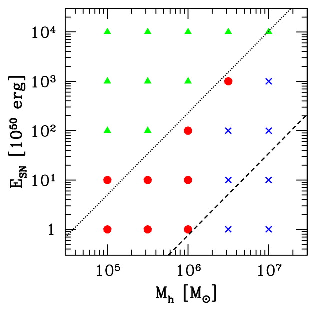}
%
%
\caption{The fate of host halos of the first stars, as a function of the mass $M_{\rm h}$ of the halo and the energy $E_{\rm SN}$ with which the stars explode as supernovae. Less massive halos have their gas blown 
away when even relatively weak (e.g. $E_{\rm SN}$ $\sim$ 10$^{50}$ erg) Pop~III supernovae explode within them, both along with ({\it circles}) and in the absence of ({\it triangles}) the additional radiative 
feedback from the progenitor stars.  Halos which retain their gas are shown by crosses. 
From equation (1) we can see that for the suite of halos shown here at $z$ $\simeq$ 20, those with $T_{\rm vir}$ $\ge$ 10$^4$ K are able to retain their gas,
in contrast to the less massive minihalos.  From Kitayama \& Yoshida (2005).}
\end{figure}

As can be inferred from this Figure, one of the distiguishing characteristics of halos massive enough to host ongoing star formation, and so to host the first galaxies, is the characteristic temperature $T_{\rm vir}$ that 
gas reaches during their virialization.  This, referred to as the virial temperature of the halo, can be derived by assuming that 
the absolute magnitude of the gravitational potential energy of the halo is twice its kinetic energy, which yields 

\begin{equation}
T_{\rm vir} \simeq 4 \times 10^4  \left(\frac{\mu}{1.2} \right) \left(\frac{M_{\rm h}}{10^8 h^{-1} {\rm M_{\odot}}} \right)^{\frac{2}{3}} \left(\frac{1+z}{10} \right)     {\rm K}   \mbox{\ ,}
\end{equation}
where $M_{\rm h}$ is the mass of the halo, $z$ is the redshift at which it collapses, and $\mu$ is the mean molecular weight of the gas in the halo, 
here normalized to a value appropriate for neutral primordial gas.  The Hubble constant $H_{\rm 0}$ = 100 $h$ km s$^{-1}$ Mpc$^{-1}$ also appears here through $h$.    
\footnote{Note that this formula is derived assuming a standard CDM cosmological model in which $h$ $\simeq$ 0.7 (see e.g. Barkana \& Loeb 2001);
as such, this formula is valid at the high redshifts (i.e. $z$ $>>$ 1) at which the first galaxies form, but must be modified at lower redshifts in order to account for a cosmological constant ${\rm \Lambda}$.}  
From Figure 1 we see that the mass of halos which are large enough to host ongoing star formation, at $z$ $\sim$ 20, is $\sim$ 10$^7$ M$_{\odot}$; this corresponds to a 
virial temperature of $T_{\rm vir}$ $\sim$ 10$^4$ K.  One of the reasons for this is that 10$^4$ K is roughly the temperature
to which photoionization by stars heats the gas (see e.g. Osterbrock \& Ferland 2006); thus, gas that is photoheated by stars remains 
bound within a halo with such a virial temperature.  In turn, the presence of this gas when stars explode as supernova leads to the rapid loss of the mechanical energy in the explosion to 
radiation, thereby limiting the amount of gas blown out of the halo, in contrast to the case of the first supernovae in less massive minihalos (see Section 3.1). 
Also, due to the efficient cooling of atomic hydrogen at this temperature, gas can collapse into halos with $T_{\rm vir}$ $\ge$ 10$^4$ K regardless of its molecular content, in 
contrast to the minihalos that host the first stars, into which primordial gas only collapses if it is cooled by H$_{\rm 2}$ molecules (e.g. Oh \& Haiman 2002);
this implies that star formation can take place even under the influence of the molecule-dissociating radiation emitted by the first stars (see Section 2.2).   

Figure 2 shows the properties of an atomic cooling halo\footnote{Because the primordial gas can cool 
via emission from atomic hydrogen and collapse into halos with $T_{\rm vir}$ $\sim$ 10$^4$ K, such halos are commonly referred to as 'atomic cooling' halos.}  
, in which a first galaxy would form, at $z$ $\sim$ 10 in a cosmological simulation (see Greif et al. 2008). 
As shown here, much of the primordial gas that falls from the intergalactic medium (IGM) 
into the potential well of the halo is shock-heated to $T_{\rm vir}$ $\sim$ 10$^4$ K at a physical distance of $\sim$ 1 kpc from the center of the halo.  
This distance corresponds to the virial radius $r_{\rm vir}$ of the halo, defined in general terms as the radius within which the average matter density is 
equal to the value at which virial equilibrium is established, which is $\simeq$ 18$\pi^2$ times the mean matter density of the universe at the redshift $z$ at which the halo forms (e.g. Barkana \& Loeb 2001).  
For the standard ${\rm \Lambda}$CDM cosmological model, this is given in physical units as

\begin{figure}[t]
\includegraphics[scale=0.515]{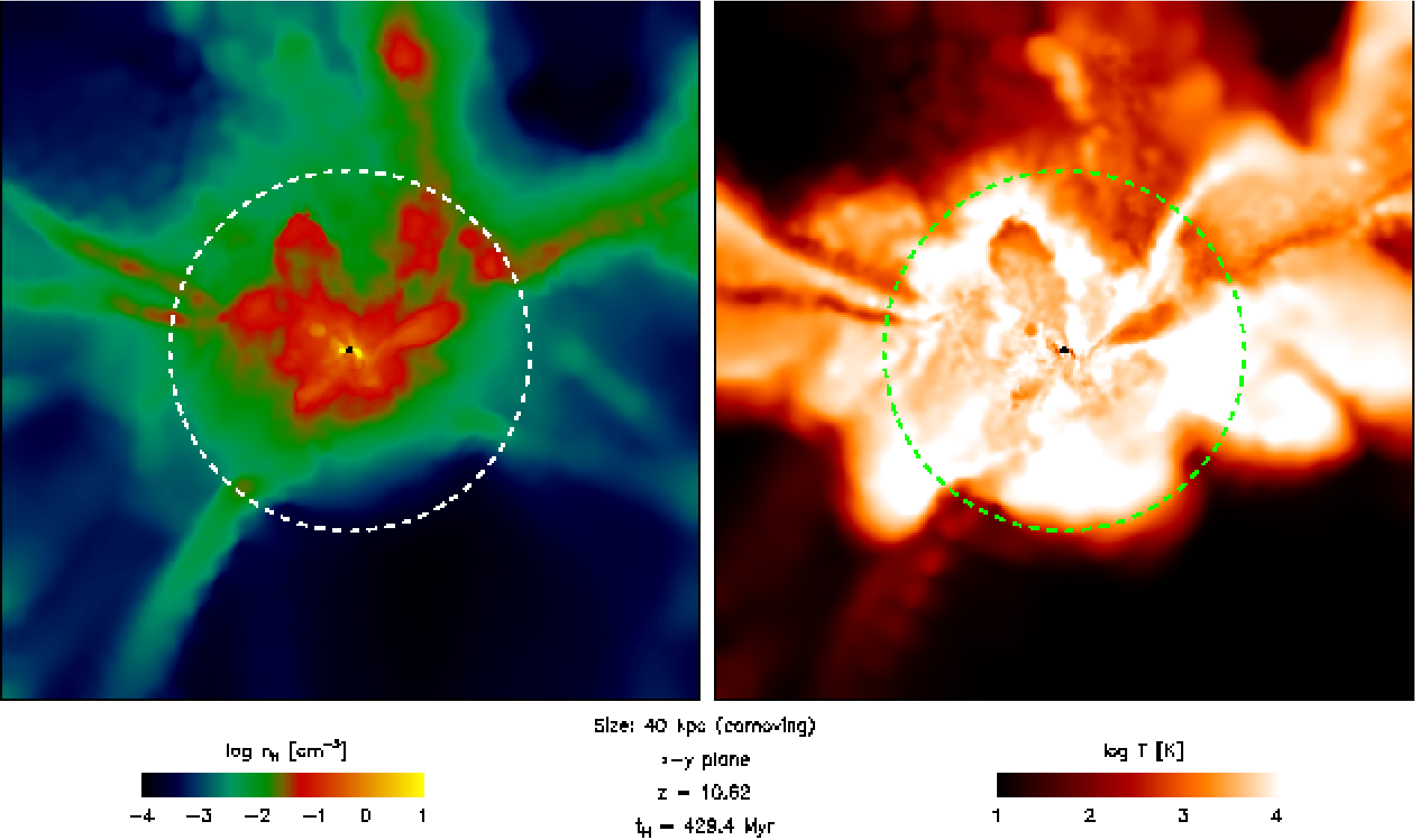}
%
%
\caption{The properties of the primordial gas collapsing into an atomic cooling dark matter halo at $z$ $\simeq $10. Shown are the hydrogen 
number density ({\it left panel}) and temperature ({\it right panel}), the dashed lines denoting the virial radius $r_{\rm vir}$ at a distance of $\simeq$ 1 kpc. Note that most of the gas is accreted 
directly from the IGM and shock-heated to the virial temperature of $T_{\rm vir}$ $\simeq$ 10$^4$ K, although cold accretion also becomes important as soon as gas cools in filaments and 
flows towards the centre of the galaxy, such as through the streams coming from the left- and right-hand sides of the panels.  In contrast to the minihalos in which the first stars form, a halo with a virial temperature $T_{\rm vir}$ $\ge$ 10$^4$ K is massive enough and has a deep enough gravitational potential well to retain its gas even when stars formed within it explode as supernovae (see Figure 1).  Hence, such halos
are strong candidates for the formation sites of the first galaxies.  From Greif et al. (2008).}
\end{figure}

\begin{equation}
r_{\rm vir} \simeq 800 h^{-1}  \left(\frac{M_{\rm h}}{10^8 h^{-1} {\rm M_{\odot}}} \right)^{\frac{1}{3}} \left(\frac{1+z}{10} \right)^{-1} {\rm pc} \mbox{\ ,}
\end{equation}
where we have normalized to values of halo mass and redshift that are typical for atomic cooling halos hosting the first galaxies.  Near the virial radius a large fraction 
of the gas is hot ($\ge$ 500 K) and rotating about the center of the halo at nearly the circular velocity $v_{\rm circ}$ of the halo (Greif et al. 2008),
defined as the velocity with which a body must move in order to be centripetally supported against gravity at the virial radius:

\begin{equation}
v_{\rm circ} = \left(\frac{GM_{\rm h}}{r_{\rm vir}} \right)^{\frac{1}{2}} \simeq 20  \left(\frac{M_{\rm h}}{10^8 h^{-1} {\rm M_{\odot}}} \right)^{\frac{1}{3}} \left(\frac{1+z}{10} \right)^{\frac{1}{2}} {\rm km \: s^{-1}} \mbox{\ .}
\end{equation} 
However, there is also a substantial portion of the infalling gas that falls to the center of the halo in cool, dense filaments and is not shock-heated to the virial temperature.  These dense
filaments feed cold gas into the central $\sim$ 100 pc of the halo, contributing to the majority of the gas the temperature of which is $<$ 500 K and which may collapse to form stars (Greif et al. 2008).

While the atomic cooling halo shown in Figure 2 is a prime example of the type of halo in which the first galaxies likely formed, there are numerous physical effects that 
were not included in the cosmological simulation from which this halo was drawn, most notably the feedback effects of Population (Pop)~III 
stars (see e.g. Wise \& Abel 2008; Johnson et al. 2008; Greif et al. 2010; Whalen et al. 2010).  
The high energy radiation emitted by the first stars both ionizes the primordial gas and dissociates molecules, which are critical cooling agents.  
Also, many of the first stars explode as violent supernovae, which inject large amounts of mechanical energy into their host minihalos and the IGM, 
as well as dispersing the first heavy elements, thereby altering forever the properties of the gas from which the first galaxies form.  
 
In this Chapter, we shall focus on how this feedback from the first generations of stars impacts the formation and evolution of the first galaxies.  In Section 2,
we briefly discuss how the cooling properties of the primordial gas, which shape the nature of Pop~III star formation, are affected by the radiation emitted from the first stars and accreting black holes.
In Section 3, we then turn to discuss how the first supernovae enrich the primordial gas with heavy elements, and how this process leads to the epoch of metal-enriched Pop~II star formation.  
In Section 4, we briefly discuss the prospects for observing the first galaxies, and for finding Pop~III star formation therein, using facilities such as the {\it James Webb Space Telescope} (JWST).
Finally, in Section 5, we close with a summary of the results presented in this Chapter and give our concluding remarks.

\section{Evolution of the Primordial Gas in the Formation of the First Galaxies}\index{Population III}
Being composed solely of the hydrogen, helium, and trace amounts of lithium and beryllium synthesized in the Big Bang, 
the primordial gas contains a limited number of coolants, chief among these H$_{\rm 2}$ at temperatures $\le$ 10$^4$ K.  Because of the ineffecient cooling 
of the gas relative to the metal-enriched\footnote{We use the common term 'metals' to refer to elements heavier than helium which are produced in stars and supernovae.} gas from which stars form today, it is likely that the Pop~III initial mass function (IMF) is top-heavy compared to that of the stars observed in our Milky Way.  
A simple explanation for this is based on the mass scale at which the fragmentation of the primordial gas takes place.  
Known as the Jeans mass $M_{\rm J}$, this is essentially the mass at which density enhancements grow via gravity more quickly than they can be erased due to pressure gradients. 
To estimate $M_{\rm J}$ for a gas with a number density $n$ and a temperature $T$, related to the sound speed $c_{\rm s}$ by 3$k_{\rm B} T$/2 = $\mu m_{\rm H} c_{\rm s}^2$/2, we first estimate
the timescale at which density enhancements grow as the free-fall time $t_{\rm ff}$ $\simeq$ ($G \rho$)$^{-\frac{1}{2}}$= ($G \mu m_{\rm H} n$)$^{-\frac{1}{2}}$,
where $G$ is Newton's constant.  Then, estimating 
the timescale in which density enhancements are erased as the sound-crossing time $t_{\rm sc}$ $\simeq$ $L/c_{\rm s}$, we equate these two timescales to estimate 
the characteristic size $L_{\rm J}$ and mass of a gas cloud which is just massive enough to collapse under its own gravity. We thus arrive at an expression for the Jeans mass $M_{\rm J}$, given by

\begin{equation}
M_{\rm J} \simeq \mu m_{\rm H} n L_{\rm J}^3 \simeq 700 \left(\frac{T}{{\rm 200 K}} \right)^{\frac{3}{2}} \left(\frac{n}{10^4 {\rm cm^{-3}}} \right)^{-\frac{1}{2}} {\rm M_{\odot}} \mbox{\ , }
\end{equation}
where we have assumed $\mu$ = 1.2, appropriate for neutral primordial gas, and have again normalized to quantities typical for primordial star-forming clouds.
As we shall discuss in Section 3, the primordial gas is in general unable to cool as efficiently as metal-enriched gas, which leads in general to higher temperatures
at fragmentation and so to a larger characteristic mass of the gravitationally unstable gas clouds from which stars form (e.g. Bromm \& Larson 2004).

While the Jeans mass is an estimate of the mass of a collapsing gas cloud, the amount of gas that is finally incorporated into a star is also dictated by the rate 
at which gas accretes onto it, starting from the formation of a protostar.  Thus, another reason that primordial stars are likely to be more massive than stars forming 
from metal-enriched gas is that higher gas temperatures also 
translate into higher accretion rates, as can be seen by estimating the accretion rate $\dot{M_{\rm acc}}$ as a function of the temperature of the gas (see e.g. Stahler et al. 1980).  Assuming 
that, through the action of gravity, the protostar grows by accreting from a gas cloud of mass $\simeq$ $M_{\rm J}$, the accretion rate can be estimated as \index{accretion}

\begin{equation}
\dot{M_{\rm acc}} \simeq \frac{M_{\rm J}}{t_{\rm ff}}  \simeq 10^{-3} \left( \frac{T}{200 {\rm K}} \right)^{\frac{3}{2}} {\rm M_{\odot}} \: {\rm yr}^{-1} \mbox{\ ,}
\end{equation}
where we have again assumed $\mu$ = 1.2 and normalized to the characteristic temperature of the gas from which Pop~III stars form in minihalos (see e.g. Glover 2005). 
Therefore, it is a combination of both the relatively large reservoir of gas available in gravitationally unstable 
gas clouds and the relatively high accretion rates onto primordial protostars (e.g. Omukai \& Palla 2003; Tan \& McKee 2004; Yoshida et al. 2008)
which suggests that Pop~III stars are more massive than metal-enriched Pop~II or Pop~I stars.  As both $M_{\rm J}$ and $\dot{M_{\rm acc}}$ depend strongly on the temperature of the gas,
one of the central questions with regard to star formation in the first galaxies is the degree to which the gas is able to cool.  In the next Section, we discuss the 
cooling of primordial gas in the first galaxies, focusing on how it is different from the case of cooling in the minihalos in which the first stars form.
Later, in Section 3.3, we discuss how the cooling properties of the primordial gas change when it is mixed with heavy elements and collapses to form the first Pop~II stars.

\begin{figure}[t]
\includegraphics[scale=0.51]{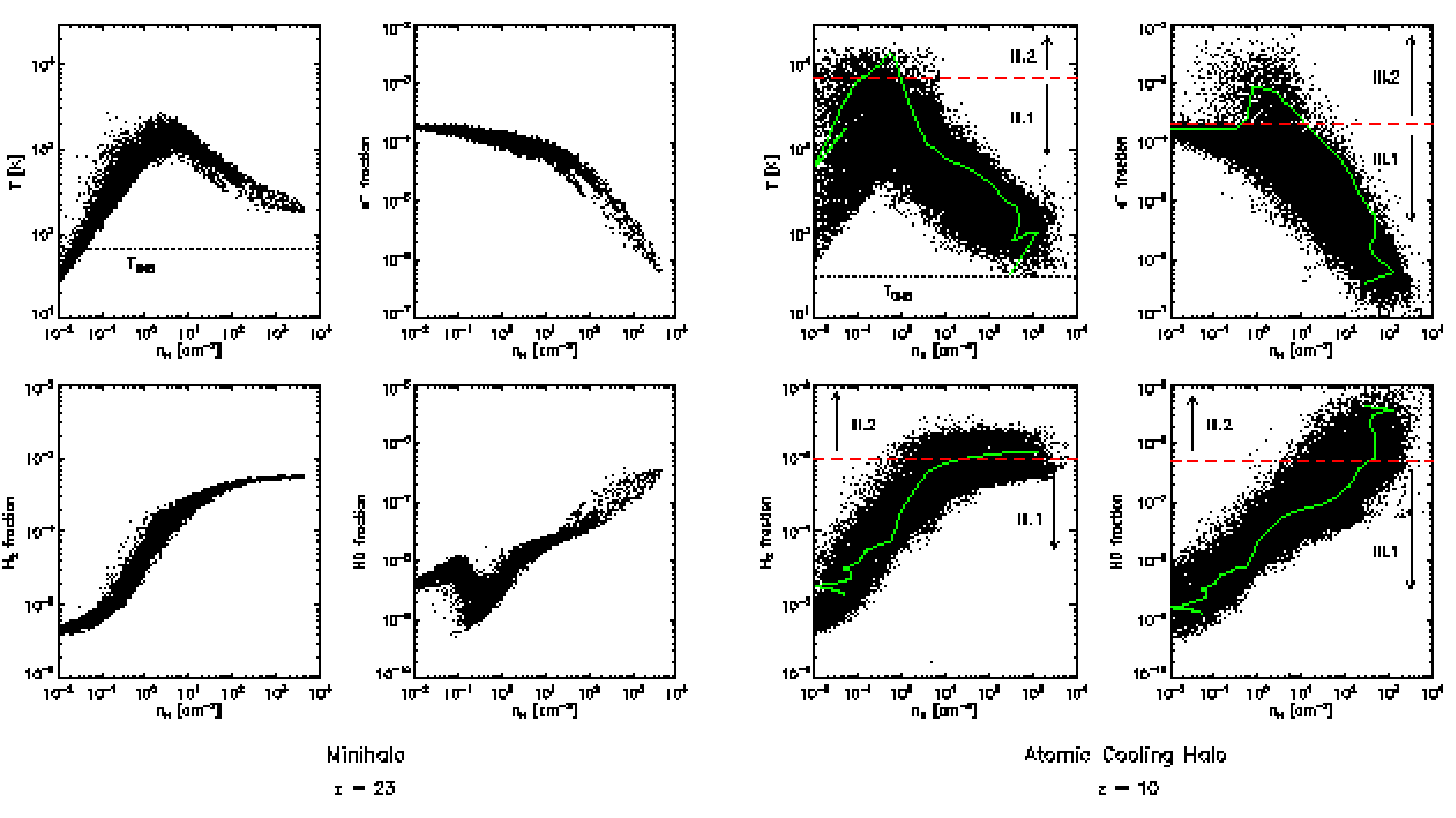}
%
%
\caption{The properties of the primordial gas inside a first star-forming minihalo at $z$ = 23 ({\it left panel}) and a first galaxy-hosting halo at $z$ = 10
({\it right panel}), as found in a cosmological hydrodynamics simulation.  The temperature, electron fraction, HD fraction, and H$_{\rm 2}$ fraction are shown as functions of number density, 
clockwise from top left to bottom left. {\it Left panel}: In the minihalo case, adiabatic collapse drives the temperature to $T$ $\ge$ 10$^3$ K and 
the density to $n$ $\ge$1 cm$^{−3}$, where molecule formation sets in and allows the gas to cool to $\simeq$ 200 K. At this point, the central clump 
becomes Jeans-unstable and presumably collapses to form at least one massive Pop~III star. {\it Right panel}: In the first galaxy, formed in an atomic cooling halo, a second cooling channel emerges 
due to an elevated electron fraction at the virial shock, where the temperature rises to $\sim$ 10$^4$ K; this, in turn, enhances molecule formation and allows the gas to cool to the much lower temperatures, as shown by 
the evolutionary track of a representative parcel of gas ({\it green lines}).  
Due to the correspondingly lower Jeans mass (equation 4) and accretion rate (equation 5), less massive Pop~III stars (Pop~III.2; see Section 2.1) are expected to form in such a first galaxy, 
perhaps with a characterisitic mass of the order of $\simeq$ 10 M$_{\odot}$. From Greif et al. (2008).}
\end{figure}

\subsection{Cooling of the Primordial Gas}\index{cooling}
As shown in Figures 2 and 3, the primordial gas collapsing into atomic cooling halos is typically shock-heated to the virial temperature of $\ge$ 10$^4$ K.  In contrast to the case of Pop~III star formation in minihalos,
the gas at these temperatures is partially ionized, and this can have important consequences for the evolution of the gas as it collapses to form stars in the first galaxies.  To see why, we note that the 
primary reaction sequence leading to the formation of H$_{\rm 2}$ molecules is (e.g. Galli \& Palla 1998; Glover 2005) \index{H$_{\rm 2}$}

\begin{equation}
{\rm e^{-}} +{\rm  H} \to {\rm H^{-}} + \gamma \mbox{\ }
\end{equation} 
       
\begin{equation}
{\rm H^{-}} +{\rm  H} \to {\rm H_{2}} + {\rm e^{-}} \mbox{\ ,}
\end{equation} 
where $\gamma$ denotes the emission of a photon.  
Whereas the primordial gas which collapses into minihalos to form the first stars has a free electron fraction $X_{\rm e}$ $\le$ 10$^{-4}$,
the collisional ionization of the primordial gas collapsing into atomic cooling halos can lead to an enhancement of the free electron fraction by a factor of more than an order of magnitude, as shown in Figure 3.
In turn, this leads to high rates of H$_{\rm 2}$ formation in atomic cooling halos, principally via the above reactions for which free electrons act as catalysts (e.g. Shapiro \& Kang 1987).  
The net result is a generally higher H$_{\rm 2}$ fraction in the high density, central regions of 
atomic cooling halos than in minihalos, as is also shown in Figure 3, and hence also to higher cooling rates due to molecular emission.   Therefore, somewhat counter-intuitively, because 
of the higher virial temperatures of the atomic cooling halos in which the first galaxies form, the dense gas in the centers of these halos can cool more effectively than in the minihalos
in which the first Pop~III stars form.   

In fact, the ionization of the primordial gas in atomic cooling halos results in the formation of another molecule which can be even more effective at cooling the gas than H$_{\rm 2}$: deutrerium hydride (HD).  
With the high H$_{\rm 2}$ fraction that develops in partially ionized gas, HD forms rapidly via the following reaction:\index{HD}

\begin{equation}
{\rm D^{+}} +{\rm  H_{2}} \to {\rm HD} + {\rm H^{+}} \mbox{\ .}
\end{equation} 
While deuterium is less abundant in the primordial gas than hydrogen by a factor of the order of 10$^{-5}$, the HD molecule is able to cool to temperatures considerably lower than
H$_{\rm 2}$ (e.g. Flower et al. 2000).  Firstly, this owes to the fact that HD has a permanent dipole moment, allowing dipole rotational transitions, which spontaneously occur much more often than the 
quadrupole rotational transitions in H$_2$.  Also, the dipole moment of HD allows transitions between rotational states of ${\rm \Delta} J = \pm 1$, which are of lower energy than the ${\rm \Delta} J = \pm 2$ quadrupole 
transitions of H$_2$.  Thus, collisions with other particles, such as neutral hydrogen, 
can excite the HD molecule from the ground to the first excited rotational state ($J$=1), from which it decays back to the ground state by a dipole transition. 
The photon that is emitted in the process carries away energy and thus cools the gas.  Because HD can be excited to the $J$ = 1 state by relatively low energy collisions, and because its 
subsequent radiative decay occurs quickly compared to that of H$_{\rm 2}$, the cooling rate per molecule is higher for HD than H$_{\rm 2}$ at temperatures $\le$ 100 K, as shown in Figure 4.

Whereas in cosmological minihalos H$_{\rm 2}$ cooling alone can cool the gas to $\simeq$ 200 K, as shown in the left panel of Figure 3, HD cooling can be so effective as to allow the primordial gas
to cool to the lowest temperature that can be achieved via radiative cooling, that of the cosmic microwave background (CMB), $T_{\rm CMB}$ = 2.7(1+$z$) (e.g. Larson 2005; Johnson \& Bromm 2006; Schneider \& Omukai 2010). 
It is useful at this point to derive this fundamental result, as we will draw on the formalism introduced here later as well, in discussing the impact of the first 
heavy elements on the cooling of the primordial gas (see Section 3.3).  

To begin, note that the frequency $\nu_{10}$ of emitted radiation for the rotational transition $J = 1 \to 0$ of HD can be expressed as 

\begin{equation}
\frac{h\nu_{10}}{k_{\rm B}} \simeq  \mbox{130 K}\mbox{\ ,}
\end{equation}
where $k_{\rm B}$ is the Boltzmann constant and $h$ is the Planck constant.  
For clarity, here we shall consider the simple case in which only this transition and its reverse occur.  
Next, consider a finite parcel of primordial gas with a temperature $T_{\rm gas}$. 
For simplicity, we shall assume that the density of the gas is sufficiently high to establish 
local thermodynamic equilibrium (LTE) level populations according to the Boltzmann distribution\footnote{Due to infrequent particle collisions at low densities,
the rate of radiative deexcitations can exceed that of collisional deexcitations, leading to non-LTE level populations (see Section 3.3).}

\begin{equation}
\frac{n_{1}}{n_{0}} =  \frac{g_{1}}{g_{0}} e^{-\frac{h \nu_{10}}{k_{\rm B} T_{\rm gas}}} \mbox{\ ,}
\end{equation}
where $n_{\rm i}$ is the number density of HD molecules in the $i$th excited rotational state and $g_{\rm i}$ is the statistical weight of that state; specifically, here we have $g_{1} = 3 g_{0}$.
Furthermore, as we are considering only transitions between the ground state and the first excited state, we shall take it that no other rotational levels are occupied.  
Equivalently, we take it here that $T_{\rm gas} < 130$ K, as otherwise collisions with other particles would be sufficiently energetic to excite the molecule to higher levels. 
Finally, we make the assumption that $T_{\rm gas} \ge T_{\rm CMB}$.  
Thus, if we denote the specific intensity of the CMB, which is an almost perfect blackbody, at the frequency $\nu_{10}$ as $I_{\nu_{10}}$, then it follows that \index{cosmic microwave background}

\begin{figure}[t]
\sidecaption[t]
\includegraphics[scale=1.05]{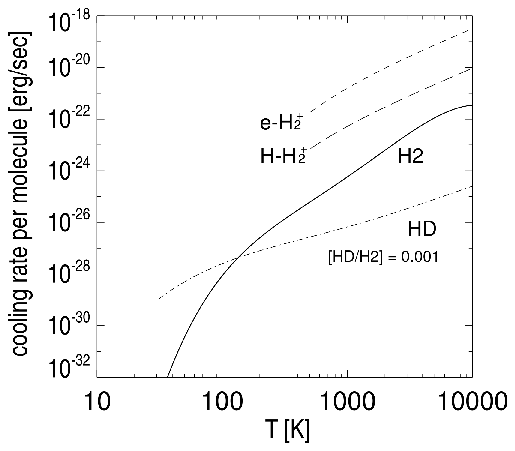}
%
%
\caption{Molecular cooling rates for H$_{\rm 2}$ ({\it solid line}), HD ({\it dot-dashed line}), and H$_2$$^+$ ({\it dashed lines}).  Owing primarily to the permanent dipole moment of HD, at low temperatures
the cooling rate per HD molecule is significantly higher than that per H$_2$ molecule.  Thus, primordial gas enriched in HD is able to cool to much lower temperatures than gas containing solely H$_{\rm 2}$; indeed, as is shown in Section 2.1, primordial gas sufficiently enriched in HD can cool to the lowest temperature attainable by radiative cooling, that of the CMB.  While the cooling rates per H$_{\rm 2}$$^+$ molecule (shown here for collisional excitation by free electrons and hydrogen atoms) can be very high, the low abundance of this molecule limits its importance for the thermal evolution of the primordial gas.  Finally, we note that at temperatures $\sim$ 10$^4$ K, the primordial gas is cooled primarily by recombination and resonance lines of atomic hydrogen (not shown).  From Yoshida et al. (2007a).}
\end{figure}

\begin{equation}
I_{\nu_{10}} = \frac{2h\nu_{10}^{3}/c^2}{e^{\frac{h\nu_{10}}{k_{\rm B} T_{\rm CMB}}}-1} \le \frac{2h\nu_{10}^{3}/c^2}{e^{\frac{h\nu_{10}}{k_{\rm B} T_{\rm gas}}}-1}\mbox{\ .}
\end{equation}   
Now, with the Einstein coefficients for spontaneous and stimulated emission from $J=1 \to 0$ denoted by $A_{10}$ and $B_{10}$, respectively, and that 
for absorption of a photon effecting the transition $J=0 \to 1$ by $B_{01}$, we have the standard relations $B_{10}g_{1} = B_{01}g_{0}$ and 

\begin{equation}
\frac{2h\nu_{10}^{3}}{c^2} = \frac{A_{10}}{B_{10}}\mbox{\ .}
\end{equation}   
Along with equations (10) and (11), these imply that


\begin{equation}
n_{0} B_{01} I_{\nu_{10}} <  n_{1}A_{10} + n_{1}B_{10}I_{\nu_{10}} \mbox{\ .}
\end{equation} 

Thus, the gas is cooled, as more energy is emitted into the CMB radiation field than is absorbed from it.  
The rate at which the temperature drops can be found by first expressing the energy density of the gas as 

\begin{equation}
u_{\rm gas} = \frac{3}{2} n k_{\rm B} T_{\rm gas} \mbox{\ ,}
\end{equation}
where $n$ is the total number density of the gas particles, including all species.  With this, equation (13) implies that, with no change in the density of the gas, 

\begin{equation}
h\nu_{10}[n_{0}B_{01}I_{\nu_{10}} -  n_{1}A_{10} - n_{1}B_{10}I_{\nu_{10}}] = \frac{3}{2}nk_{\rm B}\frac{dT_{\rm gas}}{dt} \mbox{\ .}
\end{equation}
Next, we take it that the ratio of the number density of HD molecules $n_{\rm HD}$ to the total number density of particles $n$ in the gas is given by the constant factor

\begin{equation}
X_{\rm HD} \equiv \frac{n_{\rm HD}}{n} \simeq \frac{n_{0} + n_{1}}{n} 
\simeq \frac{n_{0}}{n} \mbox{\ .}
\end{equation}        
%
%
Then using equations (10), (11), and (12) in equation (15), and neglecting stimulated emission for simplicity, the thermal evolution of the gas is approximately described by

\begin{equation}
\frac{dT_{\rm gas}}{dt} \simeq \frac{2h\nu_{10}A_{10}X_{\rm HD}}{k_{\rm B}}\left(e^{-\frac{h\nu_{10}}{k_{\rm B} T_{\rm CMB}}}-e^{-\frac{h\nu_{10}}{k_{\rm B}T_{\rm gas}}}\right)  \mbox{\ .}
\end{equation}  
It is clear from this result that if $T_{\rm CMB} \le T_{\rm gas} < 130$~K, with the gas cooling only by radiative decay of the excited rotational 
state $J$=1 to $J$=0, the temperature of the gas will asymptotically approach $T_{\rm CMB}$.  Thus, equation (17) describes the fact that the CMB 
temperature is indeed a lower limit on the temperature to which a gas can cool via line emission only.  
Using the previous equation, we can estimate the timescale for reaching the CMB temperature floor as
\begin{eqnarray}
t_{\rm CMB} & \simeq & \frac{1}{2 A_{10}X_{\rm HD}}\left(
\frac{k_{\rm B}T_{\rm CMB}}{h\nu_{10}}\right)^2
\exp\left(\frac{h\nu_{10}}{k_{\rm B}T_{\rm CMB}}\right) \nonumber \\
             & \simeq &  \left(A_{10}X_{\rm HD}\right)^{-1}\mbox{\ .} 
\end{eqnarray}

Finally, we may use this timescale to define a critical HD abundance above which the gas may cool to the CMB, by demanding 
that the gas is able to cool faster than it is heated by compression during its collapse, which takes place roughly on the free-fall timescale. We thus require that $t_{\rm CMB}\sim
t_{\rm ff}$, where the free-fall time is calculated at the characteristic density $n$ $\sim$ 10$^{4}$ cm$^{-3}$ at which the primordial gas is found to fragment in cosmological simulations (e.g. Bromm \& Larson 2004).
With $A_{10}\simeq 5\times 10^{-8}$~s$^{-1}$ for this transition (e.g. Nakamura \& Umemura 2002), we thus find the critical HD abundance to be approximately

\begin{figure}[t]
\sidecaption[t]
\includegraphics[scale=0.55]{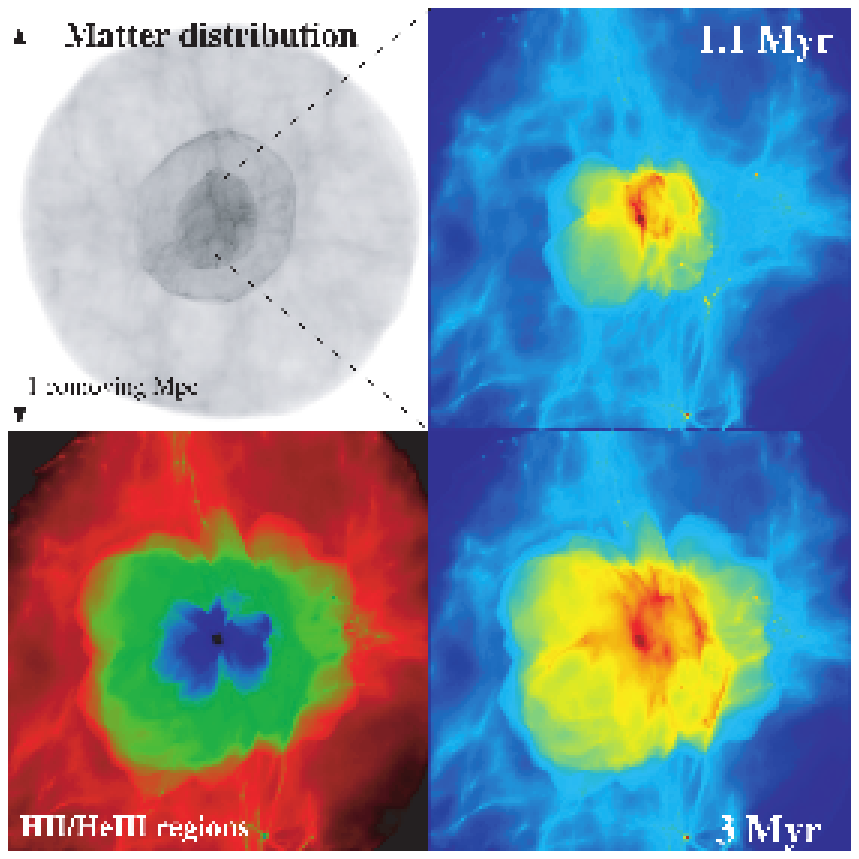}
%
%
\caption{Expansion of the H~{\sc ii} region around a Pop~III.1 star formed in a minihalo: the large-scale density distribution at $z$ 
= 26 ({\it top left}), and the projected gas density at 1.1 ({\it top right}) and 3 Myr 
({\it bottom right}) after the central star turns on. The bottom left panel shows the extent of the H~{\sc ii} region ({\it green}) and that of the He~{\sc III} 
region ({\it blue}) at 3 Myr.  When the central star turns off after this time, the strongly ionized primordial gas begins to cool and recombine, 
with H$_{\rm 2}$ and HD molecules forming in abundance.  This, in turn, enhances the cooling properties of the gas, which may later collapse and form Pop~III.2 stars in a more massive halo hosting a 
first galaxy.  From Yoshida et al. (2007a).
}
\end{figure}

\begin{equation}
X_{\rm HD, crit}\sim 10^{-6}\mbox{\ .}
\end{equation}
If the abundance of HD is lower than $X_{\rm HD,crit}$, the gas will not have time to cool to $T_{\rm gas} \simeq T_{\rm CMB}$
during its collapse.  As shown in Figure 3, for the case of the primordial gas cooling in the first galaxies a large fraction of the
gas at densities $n$ $\ge$ 10 cm$^{-3}$ has an HD abundance greater than $X_{\rm HD, crit}$, 
whereas in the case of cooling in the minihalos hosting the first stars the HD fraction is in general much lower.   

A high abundance of HD can, in general, be formed whenever the primordial gas becomes ionized.  This occurs 
through collisional ionization, as in the case of shock heating to temperatures above $\sim$ 10$^4$ K in the virialization of 
atomic cooling halos, but also occurs when the first stars formed in minihalos emit high energy radiation which photoionizes the gas.
As shown in Figure 5, a massive Pop~III star emits enough ionizing radiation to destroy almost all of the neutral hydrogen
within a distance of a few physical kiloparsec\footnote{As H$^+$ is also referred to as H~{\sc ii}, such photoionized regions formed around stars are called H~{\sc ii} regions.  Likewise, the radiation from massive stars, and especially massive Pop~III stars, can doubly ionize helium within the so-called He~{\sc iii} region (see Section 4).} (e.g. Alvarez et al. 2006; Abel et al. 2007), via the reaction

\begin{equation}
{\rm H} + \gamma \to {\rm H^+} + e^- \mbox{\ .}
\end{equation} 
Here the products are ionized hydrogen and a free electron, which has a kinetic energy equal to the energy of 
the ionizing photon minus the ionization potential of hydrogen, 13.6 eV.   This free electron is ejected
from the atom and shares it kinetic energy with other particles via collisions, thereby heating the gas to higher 
temperatures.  Typically, an equilibrium temperature of $\sim$ 10$^4$ K is established in H~{\sc ii} regions, largely 
set by a balance between the rate at which the gas is photoheated via the above reaction and the rate at which it 
is cooled by the radiative recombination and resonance emission of hydrogen (e.g. Osterbrock \& Ferland 2006).  
While in the H~{\sc ii} regions around active stars the temperature is thus too high for molecules to form in large abundances due to collisional dissociation, once the central star dies the hot ionized gas begins to cool and recombine.  Under these conditions, molecules form rapidly and a high abundance of HD can be achieved (e.g. Nagakura \& Omukai 2005; Johnson \& Bromm 2006; Yoshida et al. 2007b; McGreer \& Bryan 2008).  

Therefore, overall, primordial gas that has either been photoionized by a Pop~III star in a minihalo or which has been partially ionized during the virialization of an atomic cooling halo, may in principle collapse and cool all the way to the temperature floor set by the CMB.  This is distinct from the case of the first Pop~III star formation in minihalos (e.g. Glover 2005), and this distinction motivates the following terminology (e.g. McKee \& Tan 2008; Greif et al. 2008; Bromm et al. 2009)\footnote{Before the adoption of this terminology, Pop~III.2 was formerly referred to as Pop~II.5 in the literature (e.g. Mackey et al. 2003; Johnson \& Bromm 2006).}:

\index{Population III}
\begin{description}[Type 1]
\item[Pop~III.1]{The first generation of primordial stars formed in minihalos and not significantly affected by previous star formation.} 
\item[Pop~III.2]{Primordial stars formed under the influence of a previous generation of stars, either by the ionizing or photodissociating radiation which they emit.} 
\end{description}

Based on the enhanced cooling of the gas due to high H$_{\rm 2}$ and HD fractions, it is expected that the typical mass scale of Pop~III.2 stars
is significantly lower than that of Pop~III.1 stars (e.g. Uehara \& Inutsuka 2000; Nakamura \& Umemura 2002; Mackey et al. 2003; Machida et al. 2005; Nagakura \& Omukai 2005; Johnson \& Bromm 2006; Ripamonti 2007; Yoshida et al. 2007b).
Following equation (4), the Jeans mass for gas that cools to the temperature of the CMB, which sets a rough upper limit for the mass of Pop~III.2 
stars formed from partially ionized primordial gas, is \index{Jeans mass}

\begin{equation}
M_{\rm J} \simeq  35 \left(\frac{1+z}{10} \right)^{\frac{3}{2}} \left(\frac{n}{10^4 {\rm cm^{-3}}} \right)^{-\frac{1}{2}} {\rm M_{\odot}} \mbox{\ ,}
\end{equation}
where we have normalized to the same characteristic density at which the primordial gas fragments in Pop~III.1 star formation (e.g. Bromm \& Larson 2004).
At this fixed density, the Jeans mass is roughly an order of magnitude lower than expected for the case of Pop~III.1 star formation in minihalos.
Also due to the lower temperature of the gas in the Pop~III.2 case, the rate of accretion onto a protostar is similarly lower (Yoshida et al. 2007b):
 
\index{accretion}
\begin{equation}
\dot{M_{\rm acc}}  \simeq 5 \times 10^{-5} \left( \frac{1+z}{10} \right)^{\frac{3}{2}} {\rm M_{\odot}} \: {\rm yr}^{-1} \mbox{\ ,}
\end{equation}
compared to $\dot{M_{\rm acc}}$ $\sim$ 10$^{-3}$ M$_{\odot}$ yr$^{-1}$ for the case of Pop~III.1 stars (see equation 5).

While enhanced molecule abundances are likely to result in lower characteristic stellar masses, other mitigating effects also come into play in the 
formation of second generation primordial stars.  One factor which likely becomes important for shaping the stellar IMF, particularly in atomic cooling halos (e.g. Wise \& Abel 2007b; Greif et al. 2008),
is the development of supersonic turbulence (see e.g. Mac Low \& Klessen 2004; Clark et al. 2011b).  Also, the degree to which the abundances of H$_{\rm 2}$ and HD can be raised in the first galaxies is dependent 
on the strength of the molecule-dissociating radiation field generated by the first generations of stars (e.g. Wolcott-Green \& Haiman 2011).  In the next Section, 
we shall see that an elevated radiation field may not only result in higher Pop~III star masses, but may also result in the formation of the seeds of the first supermassive black holes.

\subsection{Suppression of Cooling by the Photodissociation of Molecules}\index{photodissociation}
The assembly of the first galaxies becomes much more complex with the formation of the first stars, in part because they emit high energy 
radiation that alters the primordial gas in dramatic ways (e.g. Ciardi \& Ferrara 2005).   As in the case of primordial gas in the minihalos in which the first stars form, in the first galaxies one of the primary 
cooling processes is the emission of radiation from molecular hydrogen, and high energy radiation emitted by the first stars can easily destroy these molecules 
(e.g. Haiman et al. 1997; Omukai \& Nishi 1999; Ciardi et al. 2000; Glover \& Brand 2001; Mackacek et al. 2001; Ricotti et al. 2001). 
So called Lyman-Werner (LW) photons, with energies 11.2 eV $\le$ $h\nu$ $\le$ 13.6 eV excite H$_{\rm 2}$, leading in turn to its dissociation into atomic hydrogen\footnote{While for simplicity we limit our discussion to the photodissociation of H$_{\rm 2}$, HD molecules are also destroyed via this mechanism.} (Stecher \& Williams 1967):

\begin{equation}
{\rm H_{2}} + \gamma \to {\rm H_{2}^{*}} \to 2{\rm H} \mbox{\ .}
\end{equation}

With the destruction of H$_{\rm 2}$ molecules, the primordial gas cools less rapidly and this signals a change in the rate at which gas can collapse into minihalos and form Pop~III stars.  An estimate of the minimum 
LW radiation field necessary to significantly delay star formation in a minihalo can be found by comparing the timescale $t_{\rm form}$ for the formation of H$_{\rm 2}$ to the timescale for its photodissociation.  
For a general radiation field the photodissociation time can be expressed as $t_{\rm diss}$ $\simeq$ 3 $\times$ 10$^4$ $J_{\rm 21}$$^{-1}$ yr, where the specific intensity $J_{\rm LW}$ of the LW radiation field 
is defined as $J_{\rm LW}$ = $J_{\rm 21}$ $\times$ 10$^{-21}$ erg s$^{-1}$ cm$^{-2}$ Hz$^{-1}$ sr$^{-1}$ (e.g. Abel et al. 1997); here, $J_{\rm 21}$ is a dimensionless parameter normalized to a typical level 
of the radiation field.  To estimate the formation time  we note that, as shown in the left panel of Figure 3, 
primordial gas collapsing into a minihalo is roughly adiabatic until its density rises to roughly $n$ $\simeq$ 1 cm$^{-3}$, at which point its temperature 
is $T$ $\simeq$ 10$^3$ K.  Therefore, it is only at this characteristic density and higher that H$_{\rm 2}$ is effective at cooling the gas, in turn leading to gravitational collapse and the formation of stars.
It is the formation time of H$_{\rm 2}$ in these conditions, which is $t_{\rm form}$ $\simeq$ 10$^6$ yr, that is to be compared to the photodissociation timescale $t_{\rm diss}$.  Equating these two timescales, 
we find a critical LW radiation field intensity of the order of $J_{\rm 21}$ $\simeq$ 10$^{-2}$, at which the suppression of H$_{\rm 2}$ formation and cooling slows the process of 
Pop~III star formation in minihalos (see Kitayama et al. 2001; Yoshida et al. 2003; Mesinger et al. 2006; Wise \& Abel 2007a; Johnson et al. 2008; Trenti \& Stiavelli 2009).  

\begin{figure}[t]
\sidecaption[t]
\includegraphics[scale=0.52]{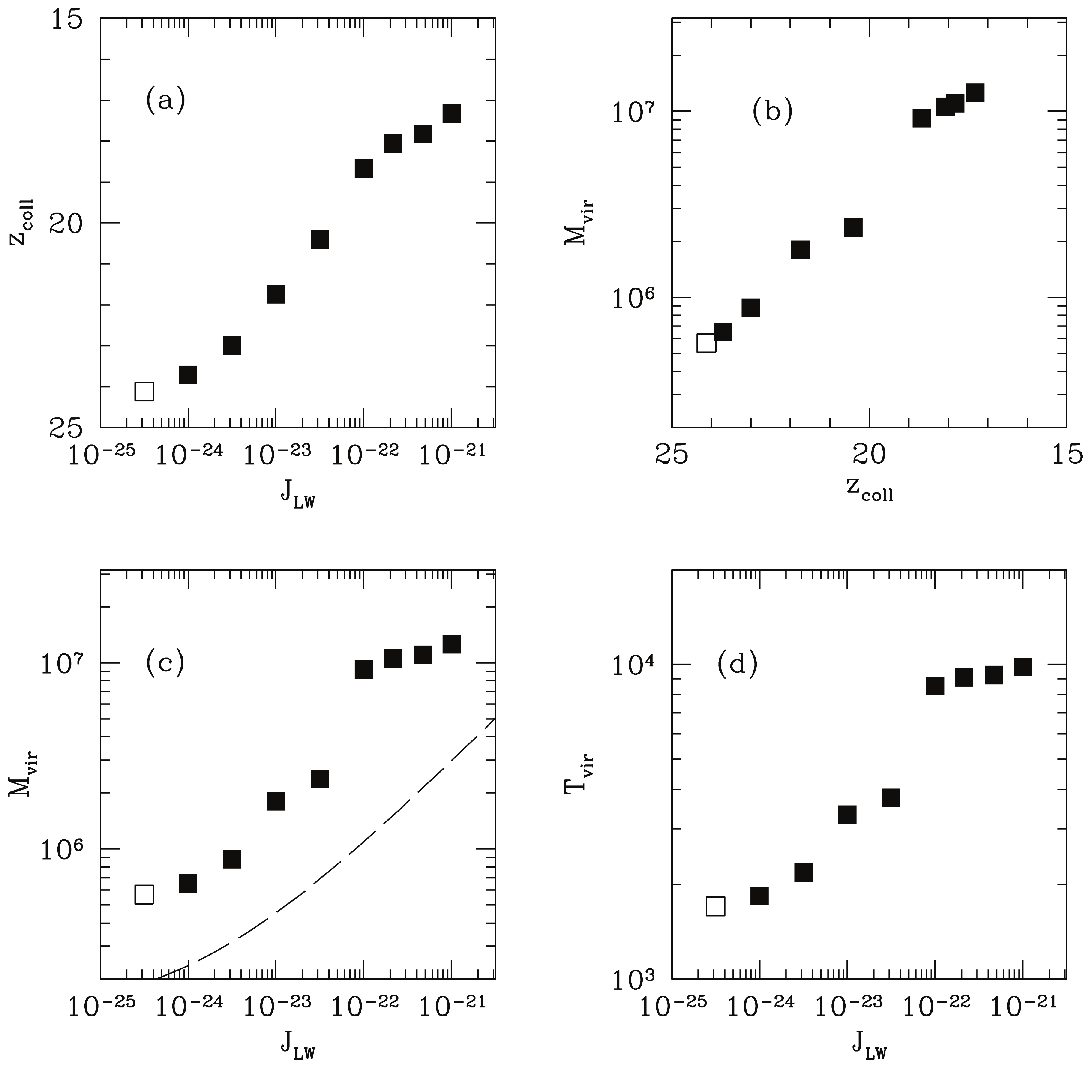}
%
%
\caption{The properties of simulated cosmological dark matter halos when the primordial gas first collapses to form Pop~III stars, 
under the influence of various levels of the background Lyman-Werner photodissociating flux $J_{\rm LW}$ (shown here in units of erg s$^{-1}$ cm$^{-2}$ Hz$^{-1}$ sr$^{-1}$):  
star formation redshift $z_{\rm coll}$ vs. $J_{\rm LW}$, $M_{\rm h}$  vs. $z_{\rm coll}$,  $T_{\rm vir}$ vs. $J_{\rm LW}$, and $M_{\rm h}$ vs. $J_{\rm LW}$ ({\it clockwise from top left panel}). 
As the intensity of the background H$_{\rm 2}$-dissociating flux increases, a given halo must grow to a larger mass and have a higher virial temperature before the primordial gas can cool, collapse
and form stars.  At a flux of $J_{\rm LW}$ $\simeq$ 4 $\times$ 10$^{-23}$ erg s$^{-1}$ cm$^{-2}$ Hz$^{-1}$ sr$^{-1}$ (i.e. $J_{\rm 21}$ $\simeq$ 0.04), there is a steep increase in the $T_{\rm vir}$ and $M_{\rm h}$ of star-forming halos due to the rate of H$_{\rm 2}$ photodissociation becoming comparable to the rate of H$_{\rm 2}$ formation.  From O'Shea \& Norman (2008).}
\end{figure}

Figure 6 shows the results of cosmological simulations of the collapse of primordial gas into minihalos, under the influence of different levels of a constant LW radiation field.  As the panels in the Figure show, 
for higher $J_{\rm LW}$ the primordial gas in a given minihalo collapses to form stars at lower redshift $z_{\rm coll}$, when the halo has grown to a higher mass $M_{\rm vir}$ and has a higher virial temperature $T_{\rm vir}$.   
The results of these simulations corroborate our estimate of the critical LW background, as $T_{\rm vir}$ and $M_{\rm vir}$ increase most dramatically at $J_{\rm 21}$ $\simeq$ 0.04.

\begin{figure}[t]
\sidecaption[t]
\includegraphics[scale=1.1]{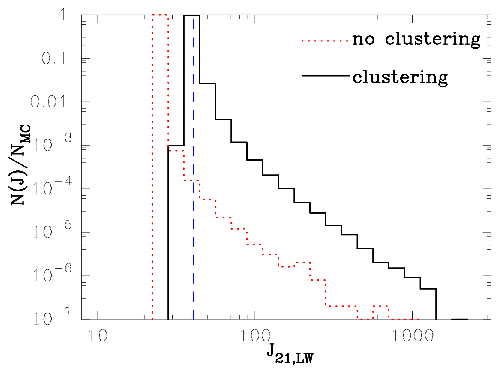}
%
%
\caption{
The probability distribution of the Lyman-Werner flux $J_{\rm LW}$ to which atomic cooling halos are exposed at $z$ = 
10, for a model accounting for the clustering of the galaxies which emit LW radiation ({\it black solid histogram}) and for 
one in which galaxies are assumed to be distributed uniformly ({\it red dotted histogram}).
The mean value of the LW background found in these models is $J_{\rm 21}$ $\simeq$ 40 ({\it dashed blue line}), which is well 
above the level needed to suppress the rate of Pop~III star formation in minihalos (see Figure 6).
While rare, a small fraction of halos are exposed to a LW flux $J_{\rm 21}$ $\ge$ 100, 
high enough to completely suppress star formation in atomic cooling halos until a $\ge$ 
10$^4$ M$_{\odot}$ black hole forms by direct collapse instead (see Shang et al. 2010).  From Dijkstra et al. (2008).}
\end{figure}

Because the mean free path of LW photons is generally large, up to $\sim$ 10 physical Mpc, a roughly uniform background field is quickly established when the first stars begin emitting radiation (e.g. Haiman et al. 1997).
We can estimate the level of the H$_{\rm 2}$-dissociating background radiation, as a function of the cosmological average star formation rate $\dot{\rho_{\rm *}}$ per unit comoving volume, by assuming that 
massive stars which live for a time $t_{\rm *}$ produce the LW flux and that $\eta_{\rm LW}$ LW photons are produced for each baryon in stars (see Greif \& Bromm 2006).  We then 
obtain for the number density $n_{\gamma}$ of H$_{\rm 2}$-dissociating photons  

\begin{equation}
n_{\gamma} \simeq \eta_{\rm LW}\frac{\dot{\rho_{\rm *}} t_{\rm *} X_{\rm H}}{m_{\rm H}}\left(1+z \right)^3 \mbox{\ ,}
\end{equation} 
where $m_{\rm H}$ is the mass of the hydrogen atom, $X_{\rm H}$ $\simeq$ 0.76 is the fraction of baryonic mass in hydrogen, and the mass density in stars is $\simeq$ $\dot{\rho_{\rm *}} t_{\rm *}$.  
Converting this to the photon energy density $u_{\gamma}$ = $h$$\nu$$n_{\gamma}$, we obtain an estimate of $J_{\rm LW}$ as a function of the star formation rate per comoving volume:

\begin{figure}[t]
\sidecaption[t]
\includegraphics[scale=1.7]{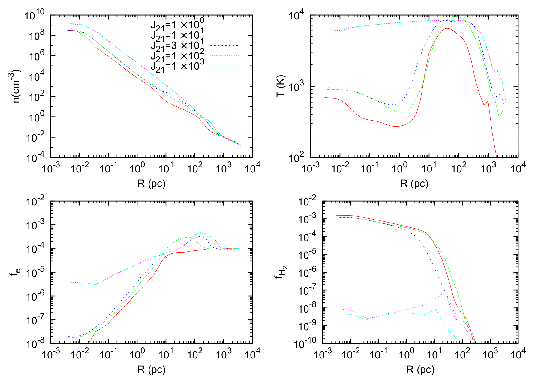}
%
%
\caption{Properties of the primordial gas in a cosmological simulation, for various elevated levels 
of the LW background radiation field $J_{\rm LW}$, here assumed to be produced by stars with an effective surface temperature 10$^4$ K, as is appropriate for metal-enriched Pop~II stars.  
Clockwise from top left: number density, temperature, H$_{\rm 2}$ fraction, and free electron fraction, each shown as a function of the distance from the center 
of an atomic cooling halo.  A LW flux with $J_{\rm 21}$ $\ge$ 100 can prevent the formation of a sufficient fraction of H$_{\rm 2}$ to
cool the gas below $\simeq$ 6 $\times$ 10$^3$ K, even at very high densities.  Due to the large Jeans mass (equation 4) and accretion rate (equation 5) 
this implies, instead of hosting Pop~III star formation, atomic cooling halos which are exposed to such an
intense LW flux are likely to host the formation of a $\ge$ 10$^4$ M$_{\odot}$ black hole by direct collapse.  From Shang et al. (2010). 
}
\end{figure}

\begin{equation}
J_{\rm LW} \simeq \frac{u_{\gamma}c}{4\pi \nu} = \frac{hc}{4\pi} \eta_{\rm LW} \frac{\dot{\rho_{\rm *}} t_{\rm *} X_{\rm H}}{m_{\rm H}} \left(1+z \right)^3  \mbox{\ , }
\end{equation}
where $c$ is the speed of light. In terms of $J_{\rm 21}$, this is

\begin{equation}
J_{\rm 21} \simeq 0.1 \left(\frac{\eta_{\rm LW}}{10^4}\right) \left(\frac{\dot{\rho_{\rm *}}}{{\rm 10^{-3} M_{\odot} yr^{-1} Mpc^{-3}}} \right) \left(\frac{1+z}{10}\right)^3
\end{equation}
where we have assumed an average lifetime $t_{\rm *}$ = 5 $\times$ 10$^6$ yr for stars that produce the bulk of H$_{\rm 2}$-dissociating radiation (e.g. Leitherer et al. 1999; Schaerer 2002).\footnote{Also note that in going from the flux estimate given by equation (25) to equation (26) we have accounted for the $\simeq$ 2.4 eV spread in LW photon energies, which is not explicitly included in equation (25).}  
For a population of metal-enriched stars formed with a Salpeter-like IMF, as is inferred for the Milky Way today, $\eta_{\rm LW}$ $\simeq$ 4 $\times$ 10$^3$; however, for metal-free stellar population with a top-heavy IMF, this can be as high as $\eta_{\rm LW}$ $\simeq$ 2 $\times$ 10$^4$ (see e.g. Greif \& Bromm 2006).  In equation (26) we have normalized to an intermediate value, for simplicity; furthermore, while the star formation rate at very high redshift is not known, we have here normalized to a rough value expected in the standard ${\rm \Lambda}$CDM picture of cosmological structure formation at $z$ $\simeq$ 10 (see e.g. Tornatore et al. 2007; Haiman 2009; Trenti \& Stiavelli 2009).

A further estimate of the cosmological background $J_{\rm LW}$, in particular that near the end of the epoch of reionization, can be found by assuming that the flux just above the Lyman limit (i.e. at $h\nu$ $\ge$ 13.6 eV) is sufficient to reionize \index{reionization} the universe (see e.g. Bromm \& Loeb 2003a; Shang et al. 2010) and that the sources producing the ionizing flux also produce a comparable flux in the LW energy range, 11.2 eV - 13.6 eV.  
Relating the number density of hydrogen nuclei at redshift $z$ to the number density $n_{\gamma}$ of ionizing photons required to keep hydrogen photoionized in the IGM, we obtain 

\begin{equation}
n_{\gamma} \simeq N_{\gamma}\frac{\Omega_{\rm b}\rho_{\rm crit}X_{\rm H}}{m_{\rm H}}\left(1+z \right)^3 \mbox{\ ,}
\end{equation} 
where $N_{\gamma}$ is the number of ionizing photons per hydrogen nucleus required to keep the universe reionized and $\Omega_{\rm b}$$\rho_{\rm crit}$ 
is the cosmological average mass density of baryons at $z$ = 0, expressed as a fraction $\Omega_{\rm b}$ of the critical density $\rho_{\rm crit}$ for a flat universe.  
Assuming that all LW photons which are emitted from sources within galaxies escape into the IGM, and taking it that only a fraction $f_{\rm esc}$ of ionizing photons are able to escape due to the 
higher optical depth to photoionization, we find an estimate of the background flux as

\begin{equation}
J_{\rm LW} \simeq \frac{1}{f_{\rm esc}} \frac{hc}{4\pi}  \frac{N_{\gamma} \Omega_{\rm b} \rho_{\rm crit} X_{\rm H}}{m_{\rm H}} \left(1+z \right)^3  \mbox{\ , }
\end{equation}
where again we have converted from photon energy density $u_{\gamma}$ = $h$$\nu$$n_{\gamma}$ to units of specific intensity as in equation (25).
Expressing this in terms of $J_{\rm 21}$, we have
\begin{equation}
J_{\rm 21} \simeq 400 \left(\frac{N_{\gamma}}{10} \right) \left(\frac{f_{\rm esc}}{0.1} \right)^{-1} \left(\frac{1+z}{10}\right)^3 \mbox{\ ,}
\end{equation}
where we have normalized $N_{\gamma}$ to the value estimated by Wyithe \& Loeb (2003), and $f_{\rm esc}$ is normalized to a 
typical value found in cosmological radiative transfer simulations (e.g. Ricotti \& Shull 2000; Ciardi \& Ferrara 2005; Wise \& Cen 2009; Razoumov \& Sommer-Larsen 2010; Yajima et al. 2011).

This estimated level of the cosmological background radiation field during reionization is well above the critical level of $J_{\rm 21}$ $\simeq$ 0.04 required 
for suppressing the rate of Pop~III star formation in minihalos, and this may have 
important implications for the nature of the stars that are formed.  In particular, under the influence of such an elevated LW background, due to the destruction of the H$_{\rm 2}$ molecules which cool the gas,
the temperature of the primordial gas when it finally collapses to form a star can be considerably higher than in the absence of a background H$_{\rm 2}$-dissociating radiation field 
(O'Shea \& Norman 2008).  This, in turn, results in a higher Jeans mass and protostellar accretion rate, likely leading to more massive Pop~III stars forming in the presence of a high LW background flux.

While the LW radiation field is in general relatively uniform, near individual galaxies it can be locally higher than the cosmological average (see Dijkstra et al. 2008; Ahn et al. 2009),
as shown in Figure 7.  In rare regions where the LW background radiation is exceptionally high, a different outcome besides Pop~III star formation in dark matter halos may result: 
the formation of a black hole by direct collapse (e.g. Bromm \& Loeb 2003a). For this to occur, the LW radiation field must be at a level high enough to destroy molecules not just 
in the outskirsts of halos where the primordial gas begins to cool via emission from H$_{\rm 2}$ molecules, but also high enough to destroy H$_{\rm 2}$ even in the central dense regions of the halo 
(but see Begelman \& Shlosman 2009; Mayer et al. 2010).  

Figure 8 shows the
results of cosmological simulations from which the minimum $J_{\rm 21}$ required for the formation of a black hole by direct collapse can be estimated.  As shown in the bottom-right panel, for $J_{\rm 21}$ $\ge$ 100 
the H$_{\rm 2}$ fraction in the gas is kept to a low level at which H$_{\rm 2}$ cooling does not lower the temperature of the gas significantly below the virial temperature of $T_{\rm vir}$ $\simeq$ 10$^4$ K of the halo
(Shang et al. 2010)\footnote{It is important to note that the spectrum of the radiation producing the LW background must also be taken into account.  While the results shown in Figure 8 are derived under the assumption 
that the LW background is generated by stars with an effective surface temperature of 10$^4$ K, appropriate for Pop~II stars, higher levels of the LW flux are required to suppress H$_{\rm 2}$ formation if, for instance,
it is generated by massive Pop~III stars with effective surface temperatures of $\simeq$ 10$^5$ K (see e.g. Shang et al. 2010).}.  
Therefore, when the gas finally collapses, the accretion rate of primordial gas will be very high, of the order of $\simeq$ 0.1 M$_{\odot}$ yr$^{-1}$, as can be seen from equation (5).  This is
roughly two orders of magnitude higher than the accretion rate onto Pop~III protostars formed in H$_{\rm 2}$-cooled gas at $T$ $\simeq$ 200 K, and the result is predicted to be 
an extremely massive 'quasi-star' which quickly collapses to form a black hole with a mass $\ge$ 10$^4$ M$_{\odot}$ (e.g. Bromm \& Loeb 2003a; Koushiappas et al. 2004; Begelman et al. 2006; Spaans \& Silk 2006; Lodato \& Natarajan 2006; Regan \& Haehnelt 2009;  et al. 2010b).  
While this level of the background LW radiation field is expected to be higher than the average, as shown in Figure 7, due to the clustering of the stars and galaxies producing LW radiation
there may be a significant number density of black holes formed by direct collapse in the early universe.  Indeed, some of these may be the seeds of the supermassive black holes observed at $z$ $\le$ 6 (see e.g. Haiman 2009).

\subsection{The Impact of Radiation from Accreting Black Holes on the Primordial Gas} \index{black holes}\index{accretion}
In addition to the radiation emitted by the first generations of stars, black holes formed and assembled into the first galaxies can also produce radiation which dramatically impacts the primordial gas.
In particular, the effects of the radiation emitted from black holes formed by direct collapse can be especially strong, as there is an initially large reservoir of gas that can be accreted onto the nascent black hole
(see Johnson et al. 2011).  To draw a comparison between the radiation emitted from stars in the first galaxies to that emitted during the accretion of gas onto black holes, we can calculate the temperature of the 
accretion disk and compare it to the typical effective temperature of a star.  For a steady accretion flow, the temperature $T$ of the accretion disk can be estimated by balancing
the rate at which the disk is heated with the rate at which it cools.  The heating is due to the gravitational potential energy of matter falling through the disk being 
dissipated by viscosity; for material falling through the disk and onto the black hole at a rate $\dot{M_{\rm BH}}$, the resultant heating rate per unit area $\Gamma$ of the disk can be estimated on dimensional
grounds as (e.g. Pringle 1981) 

\begin{equation}
\Gamma \simeq \frac{GM_{\rm BH} \dot{M_{\rm BH}}}{r^{3}} \mbox{\ .}
\end{equation}
Assuming the disk is optically thick, then the rate at which the disk cools per unit area $\Lambda$ can be estimated using the Stefan-Boltzmann law: $\Lambda$ = $\sigma_{\rm SB}$$T^4$, where 
$\sigma_{\rm{SB}}$ the Stefan-Boltzmann constant. Equating these rates yields a temperature profile for the disk. The profile thus obtained is very close to the 
following formal solution, but for a correction near the inner edge of the disk $r_{\rm in}$ where the viscous heating rate goes to zero:

\begin{eqnarray}
T(r) & = & \left(\frac{3}{8\pi}\frac{GM_{\rm{BH}}\dot{M}_{\rm{BH}}}{\sigma_{\rm{SB}} r^{3}}\right)^{\frac{1}{4}} \left[1 - \left(\frac{r_{\rm in}}{r}\right)^{-\frac{1}{2}} \right] \nonumber \\
& \simeq & 10^6 {\rm K} \left(\frac{M_{\rm BH}}{{\rm 10^4 M_{\odot}}} \right)^{-\frac{1}{4}} \left(\frac{r}{10r_{\rm s}} \right)^{-\frac{3}{4}}\mbox{\ .}
\end{eqnarray}
In the second part of the equation we have normalized to a black hole mass of 10$^4$ M$_{\odot}$, appropriate for the initial mass of a black hole formed by direct collapse.  We have also normalized the 
radius to 10 Scharzschild radii $r_{\rm s}$ = 2$GM_{\rm BH}$/$c^2$, which is well outside the inner edge of the accretion disk, $r_{\rm in}$ $\le$ 3$r_{\rm s}$. 
Finally, we have assumed accretion to take place at the Eddington rate $\dot{M_{\rm Edd}}$, at which the outward force due to electron scattering of the emitted radiation balances the inward gravitational
force acting on the accreting gas:

\begin{equation}
 \dot{M_{\rm Edd}}  =  \frac{4 \pi G M_{\rm BH} m_{\rm H}}{\epsilon c \sigma_{\rm T}} = 2 \times 10^{-4} \left(\frac{\epsilon}{0.1} \right)^{-1} \left(\frac{M_{\rm BH}}{10^4 {\rm M_{\odot}}} \right) {\rm M_{\odot} \: yr^{-1}} \mbox{\ .}
\end{equation}
Here $\sigma_{\rm T}$ = 6.65 $\times$ 10$^{-25}$ cm$^{2}$ is the Thomson cross section for the scattering of photons off electrons, and $\epsilon$ is the ratio of the radiated energy to the rest mass energy 
of the accreting material, normalized to a value appropriate for a slowly rotating black hole.  

\begin{figure}[b]
\includegraphics[scale=0.215]{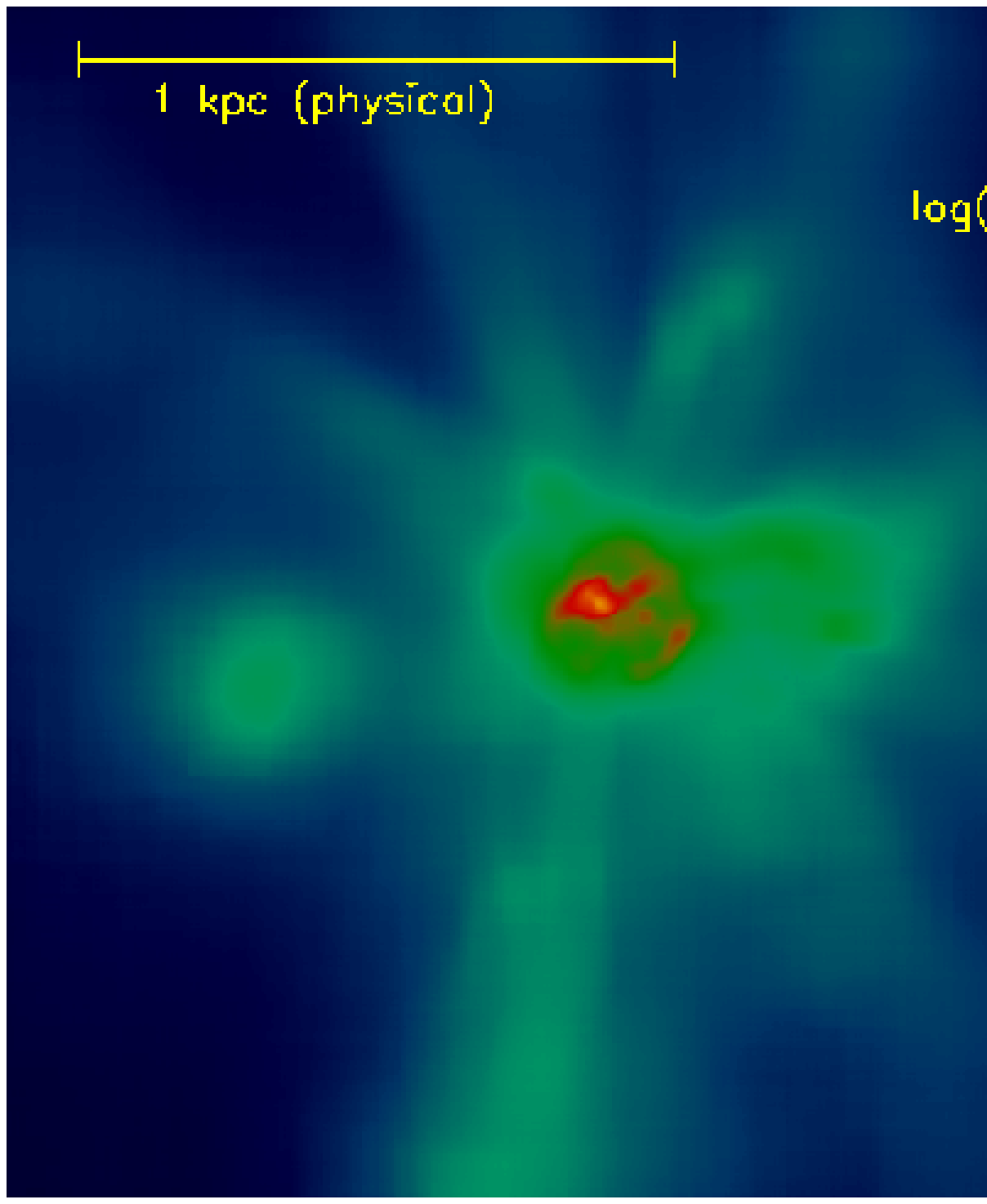}
%
%
\caption{The projected number density ({\it left panel}), temperature ({\it middle panel}), and H~{\sc ii} 
fraction ({\it right panel}) of the gas in the vicinity of an accreting 2.5 $\times$ 10$^4$ M$_{\odot}$ black hole formed by direct collapse in an atomic cooling halo at $z$ $\sim$ 12. 
The ionizing radiation emitted from the accretion disk of the black hole has a strong impact on the gas in the halo, heating it to temperatures $\simeq$ 3 $\times$ 10$^4$ K and 
causing expansion and disruption of the dense gas in the center of the halo from which the black hole feeds. From Johnson et al. (2011).}
\end{figure}

As shown in equation (31), the temperature near the inner edge of the accretion disc of a rapidly accreting black hole can be as high as $T$ $\simeq$ 10$^7$ K.
This is much higher than the effective temperature of even a very massive Pop~III star, which is roughly two orders of magnitude lower.  In turn, this implies that
accreting black holes in the first galaxies emit both copious ionizing radiation and substantial LW radiation, as well as high energy X-rays
(see e.g. Ricotti \& Ostriker 2004; Kuhlen \& Madau 2005).  In the case of a 
black hole formed by direct collapse, the resultant photoheating of the gas in the host atomic cooling halo can drive its temperature to 
 $\simeq$ 3 $\times$ 10$^4$ K, as shown in Figure 9 (Johnson et al. 2011).  Along with the associated high radiation pressure, this results in the expansion of the gas surrounding the black hole.  
The resultant drop in the density of the accreting gas translates into a decrease in the accretion rate of the black hole, which can be estimated by assuming
gas which is gravitationally bound to the black hole falls towards it at the sound speed (see e.g. Bondi 1952).  With the radius within which gas is bound to the black hole given
by $r_{\rm B}$ = $2GM_{\rm BH}/(c_{\rm s}^2+v_{\rm BH}^2)$, where $v_{\rm BH}$ is the velocity of the black hole relative to the gas, the accretion rate is estimated as the rate at which
mass passes within a distance $r_{\rm B}$ of the black hole:

\begin{eqnarray}
\dot{M_{\rm BH}} & \simeq  & \pi r_{\rm B}^2 \mu m_{\rm H} n \left(v_{\rm BH}^2 + c_{\rm s}^2 \right)^{\frac{1}{2}} = \frac{4 \pi G^2 M_{\rm BH}^2 \mu m_{\rm H} n}{(v_{\rm BH}^2 + c_{\rm s}^2)^{\frac{3}{2}}} \nonumber \\
& = & 4 \times 10^{-6} \left(\frac{M_{\rm BH}}{10^4 {\rm M}_{\odot}} \right)^2 \left(\frac{\mu}{0.6} \right)^{\frac{5}{2}} \left(\frac{n}{10^2 {\rm cm^{-3}}} \right) \left(\frac{T}{10^4 {\rm K}} \right)^{-\frac{3}{2}} {\rm M}_{\odot} \: {\rm yr}^{-1} \mbox{\ .}
\end{eqnarray}
In the second part of the equation we have assumed a black hole at rest with respect to the gas ($v_{\rm BH}$ = 0) and we have again related the gas temperature to the sound speed using  
3$k_{\rm B} T$/2 = $\mu m_{\rm H} c_{\rm s}^2$/2.
As the accretion rate is directly proportional to the density of the accreting gas and inversely proportional to its temperature, 
that the high energy radiation emitted from the accretion disk acts to heat and rarify the gas means that the accretion rate itself is regulated by the radiation generated in the process.  
Indeed, the Eddington rate given by equation (32) provides an estimate of the maximum rate at which gas can be accreted in the face of the intense radiation that is emitted. 
However, hydrodynamics calculations of accretion onto black holes formed in the first galaxies suggest that $\dot{M_{\rm BH}}$ is 
on average well below the Eddington rate because of both strong radiative feedback during accretion 
(see e.g. Pelupessy et al. 2007; Alvarez et al. 2009; Milosavljevi{\' c} et al. 2009; Park \& Ricotti 2010; Johnson et al. 2011)
and low gas densities (e.g. Yoshida 2006; Johnson \& Bromm 2007).  
This poses a challenge for the rapid growth of black holes in the early universe.

A further challenge to the model of black hole formation by direct collapse is the enrichment of the primordial gas with the first heavy elements (e.g. Omukai et al. 2008; Safranek-Shrader et al. 2010), which can easily 
cool the gas more efficiently than either H$_{\rm 2}$ or HD molecules.  We turn next to the broader question of how the first supernovae, which enrich the gas, 
transform the process of star formation in the first galaxies.

\section{Metal Enrichment and the Onset of Population II Star Formation}\index{metal enrichment}\index{Population II}
We have seen that the characteristic mass of objects that form from the runaway gravitational collapse of gas, stars and in extreme cases black holes, 
depends critically on the temperature of the collapsing gas.  The hotter the gas, the larger the Jeans mass and the higher the rate at which
gas accretes onto the collapsed object.  Therefore, the injection of heavy elements by the first supernovae represents a fundamental transition in star formation, in that 
new coolants are added to the primordial gas. As a result, the characteristic mass of stars formed from the first metal-enriched gas is likely to be lower than 
the characteristic mass of primordial stars.  Here, we investigate the transition between these two modes of star formation.

\begin{figure}[t]
\includegraphics[scale=1.06]{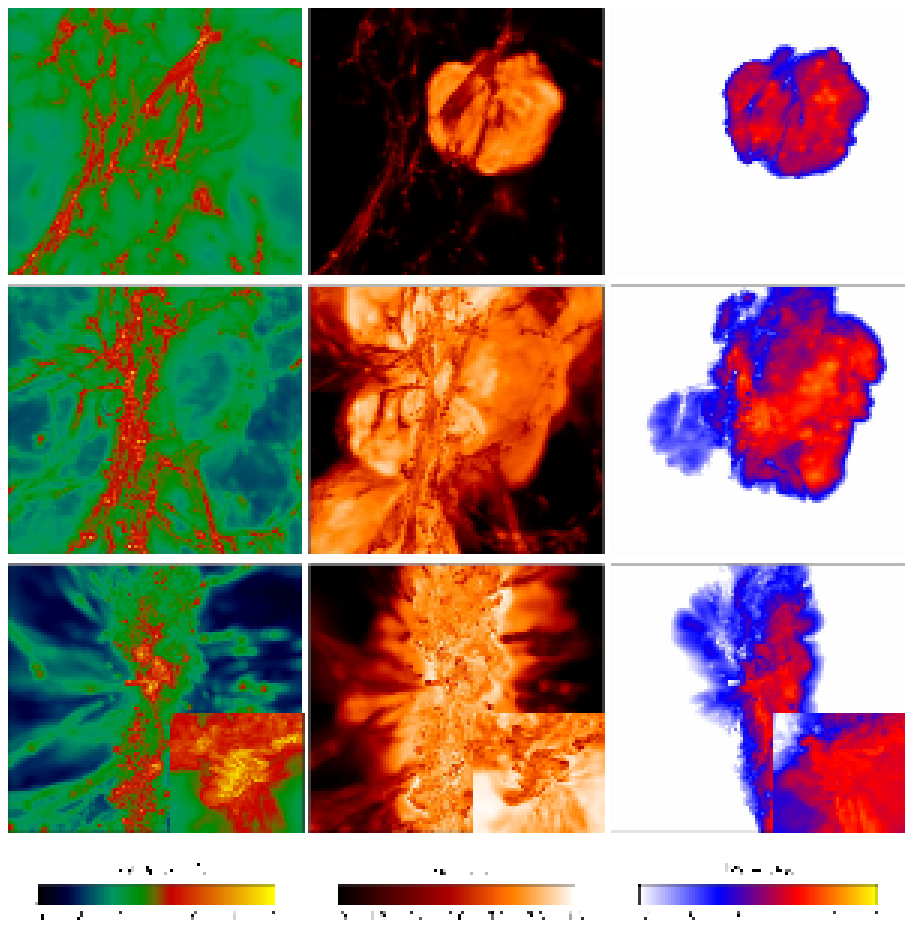}
%
%
\caption{The evolution of the remnant of a powerful Pop~III.1 supernova, exploding in a minihalo at $z$ $\sim$ 20.  
From left to right: the number density, temperature, and metallicity of the gas along the line of sight. From top to bottom: a time series showing the simulation 
15, 100, and 300 Myr after the SN explosion. Each panel is 100 comoving kpc on a side, while the inlays show the central 10 comoving kpc. 
The metals are initially distributed by the bulk motion of the SN remnant, and later by turbulent motions induced by photoheating 
from other stars and the re-collapse of the shocked gas. As can be seen in the inlays, the gas collapses to high densities once again in an atomic cooling halo
in which a first galaxy later forms from the enriched gas.  From Greif et al. (2010).}
\end{figure}

\subsection{The First Supernovae and Metal Enrichment}\index{supernovae}
It is one of the hallmark predictions of modern cosmology that the first heavy elements, such as carbon, oxygen, and iron, are produced in the cores of stars and in supernovae, rather than in the Big Bang (e.g. Burbidge et al. 1957).
  Thus, when the first stars explode as supernovae
the first metals, forged in their cores, are violently 
ejected into the primordial gas.  In this, the first supernovae introduce not just new chemical elements, but also tremendous amounts of
mechanical energy that disrupt their environments.  Indeed, as given in Section 1, the definition that we have chosen for the first galaxies pertains to this: in the first galaxies, 
formed in haloes with virial temperatures $T_{\rm vir}$ $\simeq$ 10$^4$ K, the gas can not be completely expelled by a single powerful supernova, as is the case in the 
minihalos hosting the first stars (see e.g. Bromm et al. 2003; Kitayama \& Yoshida 2005; Greif et al. 2007; Whalen et al. 2008).


The effects of a powerful Pop~III.1 supernova on the primordial gas are shown in Figure 10, as gleaned from the cosmological simulation presented in Greif et al. (2010).  
Consistent with the results presented in Figure 1, the gas within the minihalo hosting the progenitor Pop~III star is completely blown out into the surrounding IGM.  
There the primordial gas is shock-heated to several thousand Kelvin and enriched to metallicities of up to $\sim$ 10$^{-3}$ Z$_{\odot}$ within of the order of 10$^7$ yr.    
The evolution of the supernova remnant can be well described analytically, as it passes through the four distinct phases of an explosion with energy
$E_{\rm SN}$ = 10$^{52}$ erg in a medium with particle number density $n$ $\le$ 1 cm$^{-3}$, as expected for a Pop~III.1 progenitor star with a mass of the order of 100 M$_{\odot}$ 
(e.g. Fryer et al. 2001; Heger \& Woosley 2002; Whalen et al. 2008).  
At first, the blast wave from the supernova propagates outwards at a roughly constant velocity $v_{\rm sh}$; in this, the so-called free expansion phase, the distance $r_{\rm sh}$ which 
the shock has traveled from the site of the explosion by time $t_{\rm sh}$ is given simply by 

\begin{figure}[t]
\sidecaption[t]
\includegraphics[scale=1.05]{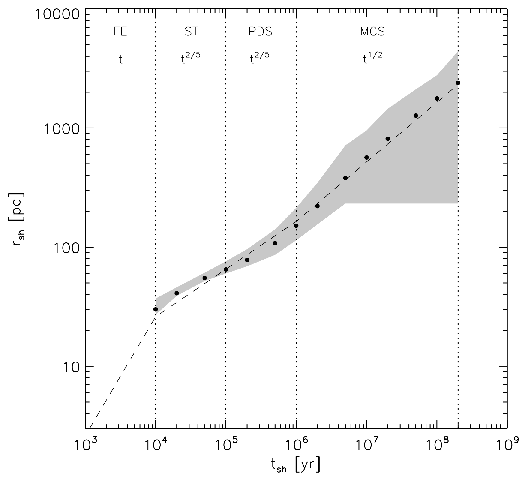}
%
%
\caption{The evolution of the remnant of a 10$^{52}$ erg Pop~III.1 supernova in a cosmological minihalo.  The analytical solution ({\it dashed lines}) discussed in Section 3.1 
accurately describes the expansion of the blast wave, the mass-weighted extent of which is shown by the black dots.  As the results shown here are from a cosmological 
simulation, the inhomogeneous density field into which the shock propagates leads to some dispersion in the distance that it extends in different directions ({\it gray region}). 
As shown in Figure 10 for a similar cosmological simulation, after of the order of 10$^8$ yr the supernova shock stalls and the metal-enriched gas recollapses into 
the growing halo which hosted the progenitor star.
From Greif et al. (2007).}
\end{figure}

\begin{equation}
r_{\rm sh} \simeq v_{\rm sh} t_{\rm sh} \simeq \left(\frac{2E_{\rm SN}}{M_{\rm ej}}\right)^{\frac{1}{2}} t_{\rm sh} \simeq 3 \left(\frac{E_{\rm SN}}{10^{52} {\rm erg}} \right)^{\frac{1}{2}} \left(\frac{M_{\rm ej}}{100 {\rm M}_{\odot}} \right)^{-\frac{1}{2}} \left(\frac{t_{\rm sh}}{10^3 {\rm yr}} \right) {\rm pc} \mbox{\ .}
\end{equation}
At this stage all of the energy of the supernova is in the kinetic energy of the ejecta, which has an initial mass $M_{\rm ej}$.
When the shock has swept up an amount of mass comparable to the original ejecta mass, the shock enters the so-called Sedov-Taylor phase in which
the energy of the blast wave is conserved while an increasing amount of mass $M_{\rm sw}$ is swept up by the shock.  In this phase we therefore have
$v_{\rm sh}$ = $dr_{\rm sh}/dt_{\rm sh}$ $\simeq$ (2$E_{\rm SN}$/$M_{\rm sw}$)$^{1/2}$, which yields for the shock radius

\begin{equation}
r_{\rm sh} \simeq 24 \left(\frac{E_{\rm SN}}{10^{52} {\rm erg}} \right)^{\frac{1}{5}} \left(\frac{n}{{\rm 1 cm^{-3}}} \right)^{-\frac{1}{5}} \left(\frac{t_{\rm sh}}{10^3 {\rm yr}} \right)^{\frac{2}{5}} {\rm pc} \mbox{\ ,}
\end{equation}
where we have used $M_{\rm sw}$ = 4$\pi$/3$r_{\rm sh}$$^3$$\mu$m$_{\rm H}$$n$, with $\mu$ = 0.6, which is appropriate for an ionized primordial gas.  The transition between
the free expansion and Sedov-Taylor phase is evident in Figure 11, which charts the propagation of the blast wave of a powerful 10$^{52}$ erg primordial supernova in a simulated cosmological minihalo,
similar to that shown in Figure 10.

The third phase, also shown in Figure 11, sets in when a substantial fraction of the original energy in the blast wave has been radiated away, principally by recombination
and resonance line cooling of the hydrogen and helium composing the primordial gas (Greif et al. 2007; Whalen et al. 2008), but also to some extent by bremsstrahlung and inverse Compton scattering of the CMB by free electrons, the latter being most important at high redshift due to the steep increase of the energy density of the CMB with redshift (e.g. Oh 2001).  Known as the pressure-driven snowplow phase, at this stage the high pressure gas behind the blast wave powers its expansion, and the equation of motion thus becomes

\begin{equation}
\frac{d(M_{\rm sw}v_{\rm sh})}{dt_{\rm sh}} = 4\pi r_{\rm sh}^2 P_{\rm b} \mbox{\ ,}
\end{equation}
where $P_{\rm b}$ is the pressure in the hot bubble interior to the blast wave.  As discussed in  Greif et al. (2007), at the radius where 
the transition to the snowplow phase begins, the density profile of the gas is close to that of an isothermal gas, $n$ $\propto$ $r_{\rm sh}^{-2}$; within this radius, the density profile is much flatter due to the 
strong photoheating of the gas by the progenitor star (e.g. Kitayama \& Yoshida 2005; and Whalen et al. 2008).  Therefore, as the pressurized bubble expands adiabatically, in the snowplow 
phase we have $M_{\rm sw}$ $\propto$ $r_{\rm sh}$ and $P_{\rm b}$ $\propto$ $r_{\rm sh}^{-5}$.  This, in turn, allows a solution to the equation of motion with $r_{\rm sh}$ $\propto$ $t_{\rm sh}^{2/5}$,
just as in the previous Sedov-Taylor phase.  

\begin{figure}[t]
\sidecaption[t]
\includegraphics[scale=0.65]{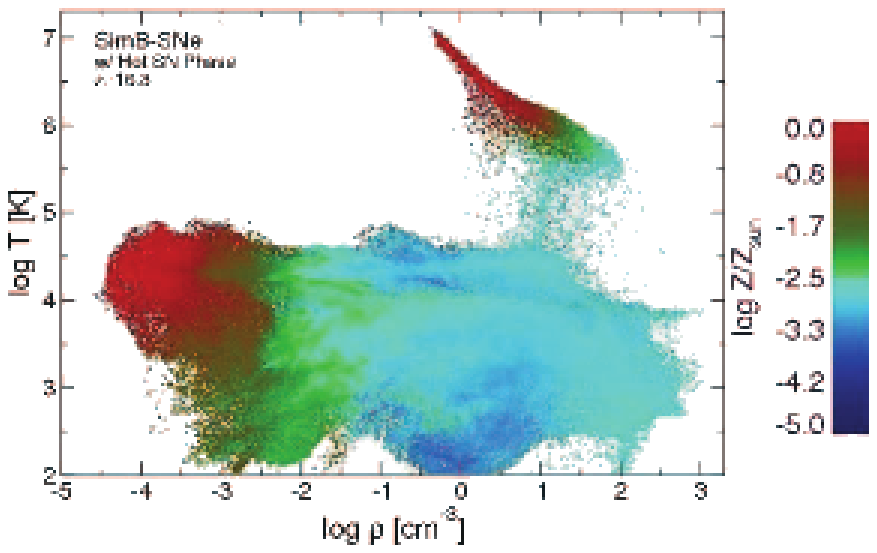}
%
%
\caption{The metallicity of the gas enriched by the violent expulsion of metals in a powerful Pop~III.1 supernova, similar
to that shown in Figures 10 and 11, as a function of the density and temperature of the gas. The highest metallicity gas 
is in the low density regions into which the blast wave propagates most rapidly.  As also shown in Figure 10, the metallicity of the higher 
density gas re-collapsing into the halo hosting the progenitor star is $Z$ $\sim$ 10$^{-3}$ $Z_{\odot}$. From Wise \& Abel (2008).}
\end{figure}

The final transition occurs when the bubble behind the blast wave has cooled and the pressure behind the shock no longer affects it dynamically.  While by this time a large fraction of the energy of the 
supernova has been radiated away, the momentum that has accumulated in the dense shell of gas that forms behind the shock is conserved.  As the density profile 
of the ambient gas is still $n$ $\propto$ $r^{-2}$ at this point, the conservation of momentum implies that the quantity $M_{\rm sw}$$v_{\rm sh}$ $\propto$ $r_{\rm sh}$ $dr_{\rm sh}$/$dt_{\rm sh}$ is a constant.  Thus, in this final phase of the supernova remnant $r_{\rm sh}$ $\propto$ $t_{\rm sh}^{1/2}$, as shown in Figure 11.

The explosion of a Pop~III.1 star with a mass of $\sim$ 200 M$_{\odot}$ as is shown Figures 10 and 11, is expected to release up to 10$^{53}$ erg as well as all
of the up to $\sim$ 100 M$_{\odot}$ in metals produced in the core of the star (e.g. Heger \& Woosley 2002; Heger et al. 2003; Karlsson et al. 2011).  
The metal-enriched gas that is ejected into the IGM by the supernova 
explosion expands preferentially into low density regions, as shown in the middle row of panels in Figure 10.  
This can be seen more explicitly in Figure 12, which shows the metallicity distribution of the gas enriched by a similar primordial supernova as a function of density and 
temperature (Wise \& Abel 2008).    

The dark matter in the halo hosting the progenitor star is not nearly so violently disrupted as is the gas swept up in the blast wave, 
and in fact the host halo continues growing until its gravity is strong enough for the cooling, metal-enriched gas to collapse into it again.  As shown in the bottom panels in Figure 10, this occurs
when the host halo has grown massive enough to host a first galaxy, as the gas is shock-heated to a temperature of $\sim$ 10$^4$ K at the virial radius $r_{\rm vir}$ $\sim$ 1 physical kpc 
from the center of the $\sim$ 10$^8$ M$_{\odot}$ halo.  
As shown in both Figures 10 and 12, the metallicity of the gas that re-collapses into the growing host halo is typically $\sim$ 10$^{-3}$ $Z_{\odot}$.  
Therefore, it is expected that stars formed in first galaxies enriched by powerful Pop~III.1 supernovae are likely enriched to this level (e.g. Karlsson et al. 2008; Greif et al. 2010; Wise et al. 2010).  
Such stars would be the first Pop~II 
stars, and as we shall see many of these stars may still be present today, 13 Gyr after the formation of the first galaxies.

\subsection{The Mixing of Metals with the Primordial Gas}\index{mixing}
Here we consider two distinct situations in which the metal-enriched ejecta of primordial supernovae mix with the primordial gas,
drawing on the results of the cosmological simluations of Pop~III supernovae discussed in Section 3.1.  Firstly, we shall estimate the timescale on which the primordial 
gas in the IGM that is swept up by the blast wave becomes mixed with the ejecta.  Then we will
turn to consider the likelihood that the primordial gas in minihalos that are overrun by the blast wave is mixed with the ejecta, thereby precluding Pop~III star formation in those halos.

When the supernova shock finally stalls after $\sim$ 10$^8$ yr, the dense shell of swept-up gas is accelerated towards the growing halo embedded in the underdense shocked
gas.  Such a configuration is Rayleigh-Taylor unstable and small perturbations of the shell can quickly grow, leading to mixing of the primordial 
gas in the shell with the metal-enriched gas in the interior.  As a stability analysis shows, a small perturbation on a length scale $\epsilon$ $<<$ $l_{\rm sh}$, 
where $l_{\rm sh}$ is the thickness of the dense shell, will grow exponentially, at a rate

\begin{equation}
\frac{d\epsilon}{dt} = \left[ \frac{2 \pi g}{l_{\rm sh}} \left(\frac{\rho_{\rm sh}-\rho_{\rm b}}{\rho_{\rm sh}+\rho_{\rm b}}\right) \right]^{\frac{1}{2}} \epsilon \mbox{\ .}
\end{equation}
Here $g$ is the acceleration of the dense shell in the direction of its interior, and $\rho_{\rm sh}$ and $\rho_{\rm b}$ are the densities of the shell
and the interior metal-enriched bubble, respectively.  Assuming that $\rho_{\rm sh}$ $>>$ $\rho_{\rm b}$, we can estimate the timescale on which the perturbation will grow as (e.g. Madau et al. 2001)

\begin{equation}
t_{\rm RT} \simeq \frac{\epsilon}{\frac{d\epsilon}{dt}} \simeq \left(\frac{2 \pi g}{l_{\rm sh}}\right)^{-\frac{1}{2}} \simeq 6 \left(\frac{l_{\rm sh}}{10 {\rm pc}} \right)^{\frac{1}{2}} {\rm Myr} \mbox{\ ,}
\end{equation}
where in the last expression the gravitational acceleration towards the growing host halo is taken to be $g$ $\simeq$ $G$$M_{\rm h}$/$r_{\rm sh}^2$, with 
$M_{\rm h}$ = 10$^8$ M$_{\odot}$, roughly the mass to which the host halo grows during the expansion of the blast wave.  
We have also used $r_{\rm sh}$ = 3 kpc, which is roughly the spatial extent of the supernova shock when it finally stalls (e.g. Greif et al. 2007).  
Even for a shell as thick as $\sim$ 100 pc, the timescale on which the metal-enriched interior material mixes with the $\sim$ 10$^5$ M$_{\odot}$ of primordial gas 
swept up by the blast wave is much shorter than the timescale on which the gas re-collapses into the host halo, which is $\sim$ 10$^8$ yr.  Therefore, the gas which re-collapses 
into the host halo is expected to be well-mixed with the metals ejected in the supernova explosion.  

\begin{figure}[t]
\includegraphics[scale=0.88]{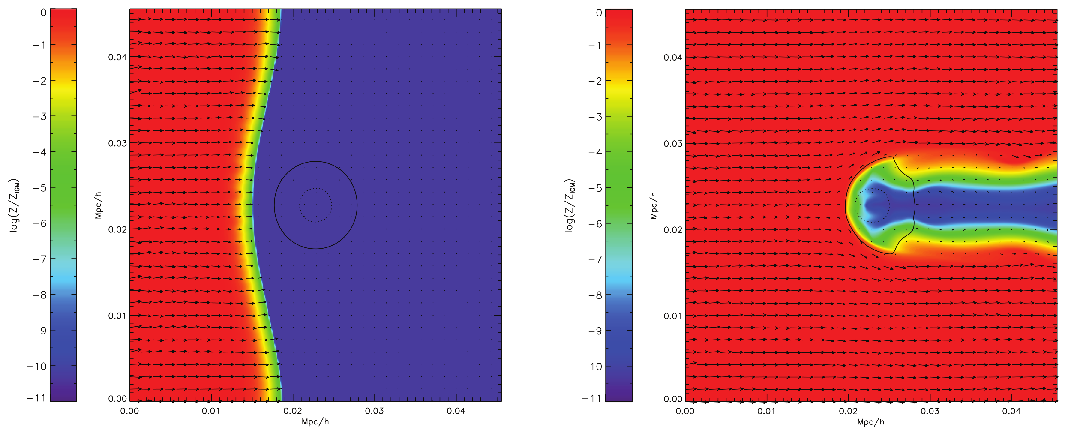}
%
%
\caption{The mixing of the primordial gas within a halo with mass $M_{\rm h}$ = 10$^7$ M$_{\odot}$ with metal-enriched gas overtaking the halo at a velocity $v_{\rm sh}$ = 30 km s$^{-1}$.  Shown is the metallicity
of the gas at $z$ = 9 ({\it left panel}) and later at $z$ = 6 ({\it right panel}).  The velocity field, along with contours of $\rho$ = $\rho_{\rm vir}$ ({\it solid lines}), is depicted in each of the panels. The primordial gas is well-mixed with the metal-enriched gas at radii $r$ $\ge$ $r_{\rm vir}$, and is mixed to a lesser degree at smaller radii, in basic agreement with equation (40).  Note that the most dense gas in the center of the halo remains pristine; as it is from this gas that stars would most likely form, such 
a halo is likely to host Pop~III star formation, despite being overtaken by high velocity metal-enriched gas.   From Cen \& Riquelme (2008).}
\end{figure}

While it is thus apparent that the low density gas swept up in the IGM can be efficiently mixed with the metal-enriched material ejected in Pop~III supernovae, 
the blast waves from these powerful explosions can also impact the more dense primordial gas inside neighboring minihalos. It is therefore another key question whether the metals are also mixed 
with this dense gas, as if so then when it collapses metal-enriched Pop~II stars may form instead of Pop~III stars.  
In this case of a supernova blast wave overtaking a dense cloud of self-gravitating gas in a minihalo, there is the possibility of the dense gas becoming Kelvin-Helmholtz unstable, in which case
vortices develop at the boundary with the fast-moving metal-enriched gas, and the two will mix with one another.  However, for this to occur the dense gas cloud must not be too tightly bound by gravity.  
In particular, for a given relative velocity between the minihalo and the blast wave, which we can take to be $v_{\rm sh}$, the gas will be mixed due to the Kelvin-Helmholtz instability at the virial radius $r_{\rm vir}$ 
of the halo, if (e.g. Murray et al. 1993; Cen \& Riquelme 2008)  

\begin{equation}
v_{\rm sh} \ge \left(\frac{g r_{\rm vir}}{2 \pi} \frac{\rho_{\rm vir}}{\rho_{\rm b}}\right)^{\frac{1}{2}} \simeq 10 \left(\frac{M_{\rm h}}{10^6 h^{-1} {\rm M_{\odot}}} \right)^{\frac{1}{3}} \left(\frac{1+z}{20} \right)^{\frac{1}{2}} {\rm km \: s^{-1}} \simeq v_{\rm circ} \mbox{\ ,}
\end{equation}
where for the second expression we have used $g$ = $G$$M_{\rm h}$/$r_{\rm vir}$$^2$ and we have assumed a density contrast between the gas at the virial radius and that of the shock $\rho_{\rm vir}$/$\rho_{\rm b}$ = 10, 
consistent with the results of the cosmological simulations of Pop~III supernvovae shown in Section 3.1. Thus, we see that the gas near the virial radius 
will be mixed if the speed of the shock exceeds the circular velocity $v_{\rm circ}$ of the halo, which is likely the case for a minihalo with mass $M_{\rm h}$ $\sim$ 
10$^6$ M$_{\odot}$ at $z$ $\le$ 20.  

However, as it is generally the dense gas embedded more deeply in the halo from which stars form, the metal-enriched gas may have to be propagating at a significantly 
higher velocity in order to impact the nature of star formation in the halo.  To estimate 
how fast the shock must be in order to mix the gas a distance $r$ from the center of the halo, we can take it that the gas in the halo 
has a density profile that is roughly isothermal, with $\rho$ $\simeq$ $\rho_{\rm vir}$ $(r/r_{\rm vir})^{-2}$.  Using the same rough scaling also for the dark matter, 
we substitute $\rho$ for $\rho_{\rm vir}$ and $r$ for $r_{\rm vir}$ in equation (39) to arrive at the following expression for the 
shock speed required for mixing via the Kelvin-Helmholtz instability:

\begin{equation}
v_{\rm sh} \ge 100 \left(\frac{M_{\rm h}}{10^6 h^{-1} {\rm M_{\odot}}} \right)^{\frac{1}{3}} \left(\frac{1+z}{20} \right)^{\frac{1}{2}} \left(\frac{r}{0.1 r_{\rm vir}}\right)^{-1} {\rm km \: s^{-1}} \mbox{\ ,} 
\end{equation}
where we have implicitly assumed the same constant $\rho_{\rm b}$ as in equation (39).
Therefore, we see that it is only relatively fast shocks that are able to efficiently mix the metal-enriched material with the dense, pristine gas in the interior of a primordial minihalo.
The results of this analysis are in basic agreement with the results of simulations of high velocity shocks impacting minihalos, as shown in Figure 13 (Cen \& Riquelme 2008).  

\begin{figure}[t]
\includegraphics[scale=0.9]{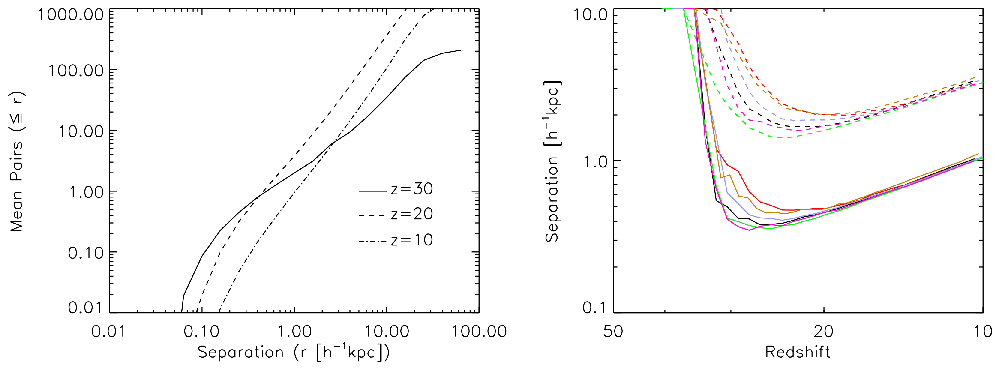}
%
%
\caption{Clustering of Pop~III star-forming minihalos in a simulation of the formation of a Milky Way-like halo. 
{\it Left panel}: mean number of progenitor halo pairs, as a function of their physical 
separation at the three redshifts indicated in the panel. {\it Right panel}: mean separation of progenitor haloes in each of 
several similar simulations. Each halo has one neighbour within the distance given by the solid line, and 10 neighbours within the distance given by 
the dashed lines. While the radiation produced by one Pop~III star in a given halo is likely to impact its neighbors and perhaps delay star formation, as 
discussed in Section 2, it is in general unlikely that the metal-enriched ejecta from Pop~III supernovae will mix with the dense star-forming
gas in neighboring halos.  From Gao et al. (2010).}
\end{figure}

To gauge the likelihood that a cosmological minihalo is indeed impacted by a Pop~III supernova shock that is sufficiently strong to enrich the material 
in its central regions, we turn to Figure 14, which shows the average distance between Pop~III star-forming 
minihalos in a cosmological simulation of the formation of a halo similar to that of the Milky Way (Gao et al. 2010).
This Figure shows that due to the clustering of such halos, the average distance between them is smaller than would be expected from a simple estimate drawn from their abundance assuming a homogeneous 
distribution.  In particular, especially at high redshifts ($z$ $\ge$ 20), the halos are closely clustered, with an average separation of roughly $\sim$ 500 pc.  From this we can estimate the average 
speed at which the blast wave from a Pop~III supernova in one minihalo impacts its nearest neighboring minihalo, using the results presented in Figure 11.  As can be seen from that Figure, the 
typical speeds with which the shock propagates at $r_{\rm sh}$ $\sim$ 500 pc from the explosion site are roughly 20-40 km s$^{-1}$.  This is high enough to disrupt the gas near the virial radius of a neighboring 
halo, but not high enough to mix the metal-enriched ejecta with the dense star-forming gas in the interior of the halo at $r$ $<$ 0.1$r_{\rm vir}$, as given by equation (40).  Indeed, a similar result is found for the case of
$v_{\rm sh}$ = 30 km s$^{-1}$ in simulations in which the mixing of the gas is resolved, as shown in Figure 13.  Therefore, we conclude that the inefficiency of mixing poses a substantial challenge for  
the metals ejected in Pop~III supernova explosions to enrich other star-forming halos and prevent Pop~III star formation from occurring (see also Wyithe \& Cen 2007; Wise \& Abel 2008; Greif et al. 2010).  

While here we have presented simple analytical estimates of the degree to which metals ejected in the first supernovae are mixed with the primordial gas via hydrodynamical instabilities, 
both in the IGM and in neighboring minihalos, other processes also contribute to mixing metals into the primordial gas (see e.g. Ferrara et al. 2000; Karlsson et al. 2011; Maio et al. 2011).  
Perhaps chief among these is the turbulence which develops 
as gas rapidly flows into the centers of the atomic cooling halos in which the first galaxies form (Wise \& Abel 2007b; Greif et al. 2008) and acts to enhance the rate at which mixing takes place on small scales 
via diffusion (see e.g. Tenorio-Tagle 1996; Klessen \& Lin 2003; Karlsson 2005; Pan \& Scalo 2007).  Once star formation begins in these halos, turbulent mixing is also facilitated by the energy injected by supernova explosions (e.g. Mori et al. 2002; Wada \& Venkatesan 2003; Vasiliev et al. 2008), and the fraction of un-enriched primordial gas in the first galaxies is expected to continually drop with time  (e.g. de Avillez \& Mac Low 2002).  We turn next to discuss the impact that the first 
metals, once mixed into the primordial gas, have on the cooling of the gas and so on the nature of star formation.

\subsection{Metal Cooling in the First Galaxies}\index{metal cooling}
In Section 2.1 we discussed how cooling by the molecule HD, which may be formed in abundance in partially ionized primordial gas, can lower the temperature 
of the gas to the lowest temperature possible via radiative cooling, that of the CMB.  Here we draw on the same formalism introduced there to show how just
a small amount of metals mixed into the primordial gas can allow it to cool to low temperatures even more efficiently.  
While a number of heavy elements contribute to the cooling of low-metallicity gas, here we shall take a simplified approach and focus only on cooling by carbon, 
which is likely to have been released in abundance in the first supernova explosions (e.g. Heger \& Woosley 2002, 2010; Tominaga et al. 2007).  

To begin, we note that once the first generations of stars
form and a background radiation field is established, as discussed in Section 2.2, 
besides dissociating H$_{\rm 2}$ this background radiation field can easily ionize neutral carbon (e.g. Bromm \& Loeb 2003b).  This makes available
the potent coolant C~{\sc ii} which, even in the Galaxy today, is important for cooling the gas to very low temperatures in dense star-forming clouds (e.g. Stahler \& Palla 2004).  
\begin{figure}[t]
\includegraphics[scale=0.92]{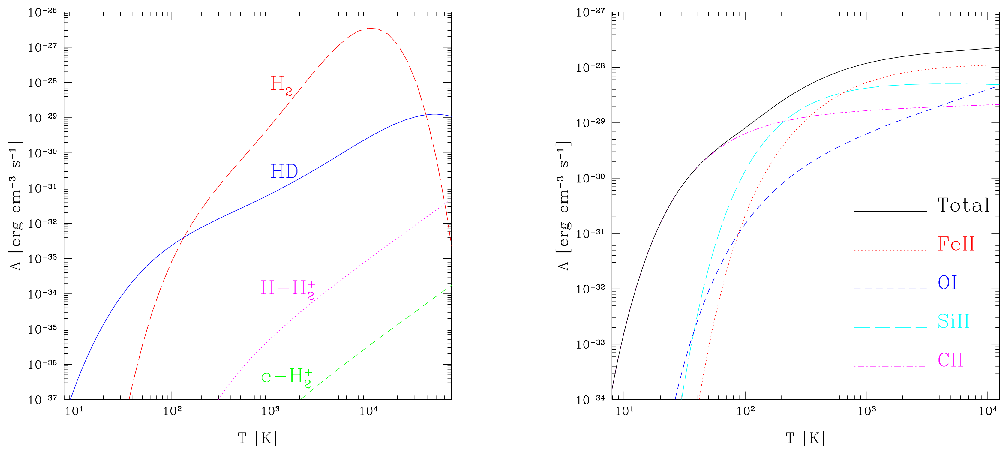}
%
%
\caption{{\it Left panel}: Cooling rates for primordial gas with a hydrogen number 
density $n_{\rm H}$ = 1 cm$^{-3}$ and the following fractions for the different species: 
X$_{\rm HD}$ = 10$^{-8}$ , X$_{\rm H_2}$ = 10$^{-5}$ , X$_{\rm H_2^+}$ = 10$^{-13}$ , X$_{\rm e^-}$ = 10$^{-4}$ . The H$_{\rm 2}$ 
cooling rate ({\it long-dashed line}) is plotted together with those of HD ({\it solid}), 
H-impact H$_{\rm 2}$$^+$ ({\it dotted line}) and e-impact H$_{\rm 2}$$^+$ ({\it short-dashed line}). 
{\it Right panel}: Cooling rates of various metal species as a function of temperature, 
for a gas also with hydrogen number density $n_{\rm H}$ = 1 cm$^{-3}$; for each 
metal species a fractional abundance of 10$^{-6}$ is assumed. The cooling rate per C~{\sc ii} ion is higher 
than the cooling rate per particle of any of the other species shown here, at $T_{\rm gas}$ $\le$ 100 K.  From Maio et al. (2007). 
}
\end{figure}
To see how the presence of C~{\sc ii} in the first galaxies affects the cooling of the gas, we first note that at low temperatures this ion 
is readily collisionally excited from its ground $^2P_{\rm 1/2}$ state to the first excited $^2P_{\rm 3/2}$ state.  The energy
difference between these two fine-structure states is just

\begin{equation}
\frac{h\nu_{10}}{k_{\rm B}} \simeq  \mbox{92 K}\mbox{\ ,}
\end{equation}
where here $\nu_{\rm 10}$ denotes the frequency of the photon emitted in the radiative decay of the first excited state back to the ground state.  As the energy difference is even smaller than
that between the ground and first excited rotational states of HD, C~{\sc ii} offers the potential to more efficiently cool the gas than HD, as even 
lower energy collisions are able to excite the ion.  In addition, the Einstein coefficient for spontaneous radiative decay is $A_{\rm 10}$ = 2.4 $\times$ 10$^{-6}$ s$^{-1}$,
almost two orders of magnitude higher than that for the $J$ = 1 $\to$ 0 transition of HD.  

We can obtain a conservative lower limit for the cooling rate of C~{\sc ii} via this transition by considering the cooling of 
gas in the low density regime, in which the rate of collisional excitations is balanced by the rate of radiative decays, that is at densities $n$ $<$ $n_{\rm crit}$, where the critical density $n_{\rm crit}$ is defined
as that above which the rate of collisional deexcitations exceeds the rate of radiative deexcitations.  For the transition of C~{\sc ii} that we are considering, $n_{\rm crit}$ = 3 $\times$ 10$^3$ cm$^{-3}$.  In this case,
the cooling rate is given as (e.g. Stahler \& Palla 2004)

\begin{equation}
\Lambda_{\rm CII}(n < n_{\rm crit}) \simeq \frac{g_{\rm 1}}{g_{\rm 0}}n_{\rm 0}n_{\rm H}\gamma_{\rm 10}h\nu_{\rm 10}e^{\frac{-h\nu_{\rm 10}}{k_{\rm B}T_{\rm gas}}} =1.5 \times 10^{-23} \left(\frac{X_{\rm CII}}{10^{-6}}\right) \left(\frac{n_{\rm H}}{10^3 {\rm cm^{-3}}}\right)^2 e^{-\frac{92 {\rm K}}{T_{\rm gas}}} {\rm erg \: s^{-1} \: cm^{-3}} \mbox{\ ,}
\end{equation}
where $\gamma_{\rm 10}$$n_{\rm H}$ = 6 $\times$ 10$^{-10}$ $n_{\rm H}$ s$^{-1}$ is the rate at which a given C~{\sc ii} ion in the ground state is excited due to a collision with a neutral hydrogen atom, $g_{\rm i}$ 
is the statistical weight of the $i$th excited state, $X_{\rm CII}$ is the fractional abundance of C~{\sc ii} relative to hydrogen, and $T_{\rm gas}$ is the temperature of the gas.  
For densities $n$ $>$ $n_{\rm crit}$, in turn, the cooling rate varies linearly with the density of the gas, since in this case the level populations are in LTE, as given by equation (10), and the rate of radiative
decay is no longer balanced by the rate of collisional excitation.  In this case, we have

\begin{equation}
\Lambda_{\rm CII}(n > n_{\rm crit}) \simeq \frac{g_{\rm 1}}{g_{\rm 0}}n_{\rm 0}A_{\rm 10}h\nu_{\rm 10}e^{\frac{-h\nu_{\rm 10}}{k_{\rm B}T_{\rm gas}}} = 6 \times 10^{-22} \left(\frac{X_{\rm CII}}{10^{-6}}\right) \left(\frac{n_{\rm H}}{10^4 {\rm cm^{-3}}}\right) e^{-\frac{92 {\rm K}}{T_{\rm gas}}} {\rm erg} \: {\rm s^{-1}} \: {\rm cm^{-3}} \mbox{\ .}
\end{equation}

The right panel of Figure 15 shows, along with the cooling rates of a number of other metal species, the cooling rate due to C~{\sc ii} given above, as a function of gas temperature, for $n$ $<$ $n_{\rm crit}$.  
The cooling rates of oxygen, iron, and silicon that are shown can be obtained following the first part of equation (42) using the atomic data corresponding to those elements 
(see e.g. Santoro \& Shull 2006; Maio et al. 2007).  We note, however, that 
the cooling rate per C~{\sc ii} ion is higher than that of any of the other metal species shown, as well as being at least an order of magnitude higher than the 
cooling rate per molecule of any of the primordial species shown in the left panel, at temperatures $T_{\rm gas}$ $\le$ 100 K.  Therefore, we can focus on this 
chemical species as a means to derive a simple estimate of the minimum abundance of heavy elements required to significantly alter the cooling properties of the primordial 
gas, and perhaps thereby alter the nature of star formation.

A rough estimate of the minimum carbon abundance required for the characteristic fragmentation mass to change from the relatively large value expected for primordial gas 
in the case of Pop~III.1 star formation can be found by considering the properties of the primordial gas when fragmentation takes place.  At this stage,
the so-called loitering phase in the collapse of the primordial gas in minihalos, $T_{\rm gas}$ $\sim$ 200 K and $n$ $\sim$ 10$^4$ cm$^{-3}$ (e.g. Abel et al. 2002; Bromm et al. 2002).
Hence, the Jeans mass (equation 4)  is of the order of 100 M$_{\odot}$ and, if the gas does not cool efficiently then 
a massive Pop~III star, or perhaps a binary or small multiple system, will likely form (e.g. Turk et al. 2009; Stacy et al. 2010; Clark et al. 2011a; Greif et al. 2011).  However, 
if the gas cools to lower temperatures, then the Jeans mass becomes smaller and the gas is expected to fragment into smaller clumps; in turn, this is expected to translate 
into the formation of less massive stars.  

\begin{figure}[t]
\sidecaption[t]
\includegraphics[scale=0.65]{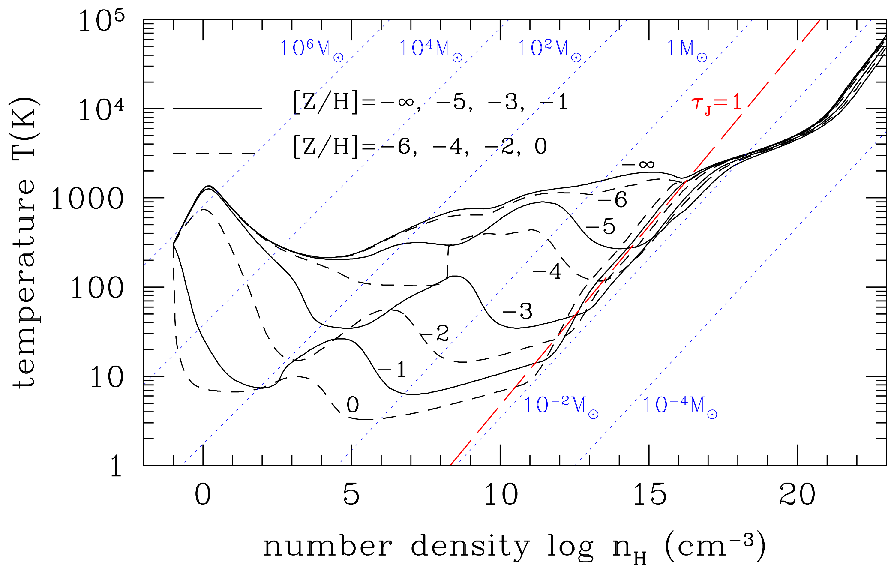}
%
%
\caption{Temperature evolution of collapsing prestellar clouds with different metallicities, as a function of density. The evolutionary tracks of clouds 
with metallicities [Z/H]=-$\infty$ (Z=0), -5, -3, and -1 (-6, -4, -2, and 0) 
are shown by solid (dashed) lines. The dotted lines denote various values of the Jeans mass $M_{\rm J}$. 
Near the critical density of H$_{\rm 2}$, $n$ $\sim$ 10$^4$ cm$^{-3}$, at which the cooling of primordial gas becomes less efficient, 
the temperature continues to drop as the gas collapses to higher densities for metallicities $Z$ $\ge$ 10$^{-4}$ $Z_{\odot}$; this is
due to the cooling provided by C~{\sc ii} and other atomic species, as discussed in the text.  At higher densities, $n$ $\ge$ 10$^{10}$ cm$^{-3}$,
the gas is able to cool efficiently even at a metallicity of $Z$ $\sim$ 10$^{-5}$ $Z_{\odot}$, if dust is present, as is the case in the calculation shown here.  From Omukai et al. (2005).}
\end{figure}

Following the discussion in Section 2.1, we note that 
in order for the gas to cool efficiently at this stage, the cooling rate must exceed the rate at which the gas is heated adiabatically by compression during its collapse (Bromm \& Loeb 2003b).  Taking the 
adiabatic heating rate to be $\Gamma_{\rm ad}$ $\sim$ 1.5$n$$k_{\rm B}$$T_{\rm gas}$/$t_{\rm ff}$, where $t_{\rm ff}$ $\simeq$ ($G \rho$)$^{-\frac{1}{2}}$ is the free-fall time and the cooling rate $\Lambda_{\rm CII}$ is 
given by equation (43), this condition is satisfied if $X_{\rm CII}$ $>$ 7 $\times$ 10$^{-8}$.  Assuming that all carbon is in the form of C~{\sc ii} and taking it that the solar abundance of carbon is
$\sim$ 3 $\times$ 10$^{-4}$ by number, this yields a critical carbon abundance of [C/H]$_{\rm crit}$ $\simeq$ -3.5.\footnote{Here we use the common notation for abundance ratios relative to those of the sun given by
[X/Y] = log$_{\rm 10}$($N_{\rm X}$/$N_{\rm Y}$) - log$_{\rm 10}$($N_{\rm X}$/$N_{\rm Y}$)$_{\odot}$, where $N_{\rm X}$ and $N_{\rm Y}$ are the numbers of nuclei of elements X and Y, respectively.}

While other elements, such as oxygen, iron, and silicon, also contribute to the cooling of metal-enriched gas, this abundance of carbon relative to the solar value is
similar to what is found for the overall critical metallicity $Z_{\rm crit}$ / $Z_{\odot}$ $\sim$ 10$^{-3.5}$ that is typically found in detailed calculations including atomic cooling\footnote{As the cooling rates of the various atomic species each contribute separately 
to the total cooling rate, it is the combination of their individual abundances which determines whether the 'critical metallicity' is achieved (see e.g. Frebel et al. 2007).} 
(e.g. Bromm et al. 2001; Omukai et al. 2005; Smith \& Sigurdsson 2007; Smith et al. 2009; Aykutalp \& Spaans 2011; but see Jappsen et al. 2009a,b).  Figure 16 shows the results of one such calculation, in which the temperature evolution of the gas is modeled as it collapses to high densities,
for various values of the metallicity of the gas.  For the case of metal-free gas, the temperature of the gas increases after the loitering phase at $n$ $\sim$ 10$^{4}$ cm$^{-3}$; in this case,
fragmentation at mass scales smaller than of the order of 100 M$_{\odot}$ is thus unlikely.  However, when $Z$ $\ge$ 10$^{-4}$ $Z_{\odot}$, close the value we found above for the critical carbon abundance, 
the gas cools as it collapses to densities $n$ $>$ 10$^{4}$ cm$^{-3}$ and consequently the fragmentation scale decreases appreciably compared to the primordial case. Hence, less 
massive stars are likely to form in gas enriched to this level.  

\begin{figure}[t]
\includegraphics[scale=0.43]{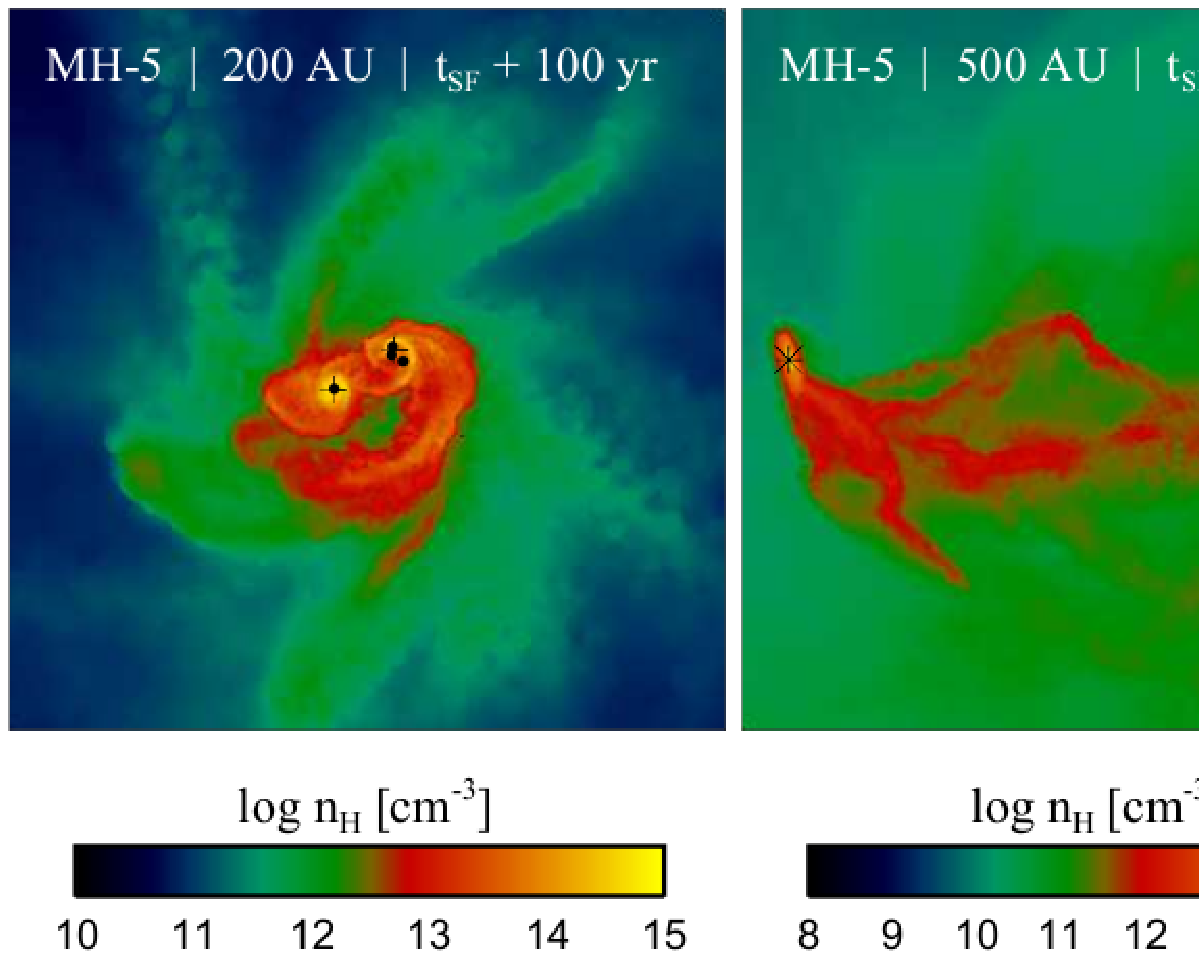}
%
%
\caption{The state of the central gas cloud in a primordial minihalo at 100 yr ({\it left panel}), 300 yr ({\it middle panel}), and 10$^3$ yr ({\it right panel}) after the formation of the first protostar
in a very high resolution cosmological simulation. 
Shown here is the hydrogen density projected along the line of sight. Blackdots, crosses and stars 
denote protostars with masses below 1M$_{\odot}$, between 1M$_{\odot}$ and 3M$_{\odot}$, and above 3M$_{\odot}$, respectively. The gas fragments into a 
relatively rich cluster of protostars with a range of masses. Dynamical interactions can lead to the ejection of low-mass protostars, while 
more massive protostars tend to remain at the center of the cloud and continue to accrete from the surrounding envelope of gas. 
Thus, while massive primordial stars are likely to form in such clusters, some fraction of Pop~III stars with masses below $\sim$ 1 M$_{\odot}$ may also form.
From Greif et al. (2011).}
\end{figure}

We note also a second drop in the temperature of the gas at higher densities for even lower metallicities in Figure 16; this decrease in temperature for $Z$ $\ge$ 10$^{-5}$ $Z_{\odot}$ occurs because of dust cooling.  
While the dust fraction in extremely metal-poor gas is not known, dust formation in early supernovae (e.g. Nozawa et al. 2003; Schneider et al. 2004; Cherchneff \& Dwek 2010) 
may yield it high enough for the thermal evolution of the gas to be affected at $n$ $\ge$ 10$^{10}$ cm$^{-3}$, as shown here, even for such extremely low metallicities.  
In this case, the critical metallicity for low-mass star 
formation may be smaller than we estimated above for the case of cooling by atomic species such as C~{\sc ii}, perhaps as low as $Z_{\rm crit}$ $\sim$ $10^{-5}$ $Z_{\odot}$ (see e.g. Schneider et al. 2006; Clark et al. 2008).

In general, simulations of the evolution of low-metallicity star-forming gas give the same general result that the fragmentation scale, as well as the protostellar accretion rate, 
is higher for metal-free gas than for metal-enriched gas, and hence that the typical masses of Pop~III stars are higher than those of Pop~II stars (e.g. Bromm et al. 2001; Smith et al. 2007; 
but see Jappsen et al. 2009a,b).  However, as shown in Figure 17, recent very high resolution cosmological simulations suggest that low-mass protostars formed in clusters 
may be ejected from the dense central regions of primordial minihalos due to dynamical interactions, 
in which case their growth may be limited due to the accretion of gas being dramatically slowed (Greif et al. 2011; see also Clark et al. 2011a).  In this event, it is possible that low-mass 
stars may indeed form from primordial gas, although they may only constitute a small fraction of all Pop~III stars (e.g. Tumlinson et al. 2006; Madau et al. 2008).  
If their masses were less than $\simeq$ 0.8 M$_{\odot}$, then such low-mass stars could be detectable
as un-enriched dwarfs or red giants in the Galaxy even today (Johnson \& Khochfar 2011), although there is a strong possibility that their surfaces would be enriched 
due to accretion of metals from the interstellar medium (see e.g. Suda et al. 2004; Frebel et al. 2009; Komiya et al. 2010).  
To date, however, no low-mass stars with overall metallicity below of the order of 10$^{-4}$ $Z_{\odot}$ have been detected, which is consistent with the critical metallicity
being set by cooling due to atomic species such as carbon and oxygen (see Frebel et al. 2007).  

While the additional avenues for radiative cooling provided by even trace amounts of metals clearly alter the evolution of the gas and the process of star formation, 
other factors also play a role in dictating the thermal and dynamical state of the gas in the first galaxies.  Magnetic fields 
may impede the large-scale collapse of the gas into dark matter halos (e.g. Schleicher et al. 2009; Rodrigues et al. 2010; de Souza et al. 2011) or alter the collapse of the gas at smaller scales during star formation
 (e.g. Kulsrud et al. 1997; Silk \& Langer 2006; Xu et al. 2008; Schleicher et al. 2010a). \index{magnetic fields}  Also, cosmic rays generated in the first supernova explosions are 
an additional source of ionization that can speed the formation of molecules and so enhance the cooling of the gas (see Vasiliev \& Shchekinov 2006; Stacy \& Bromm 2007; Jasche et al. 2007).
Finally, the impact of the turbulence generated by both the accretion of gas from the IGM and supernovae in the first galaxies
may dramatically impact the process of star formation, in general acting to decrease the mass scale at which the gas fragments and forms stars (e.g. Padoan et al. 2007; Clark et al. 2008, 2011b; Prieto et al. 2011).

\section{Observational Predictions and the Outlook for Identifying the First Galaxies}\index{Population III}
While the enrichment of the primordial gas by metals ejected in the first supernovae is likely to preclude primordial star formation in 
a large fraction of the first galaxies (Johnson et al. 2008; Wise \& Abel 2008; Greif et al. 2010; Maio et al. 2010), it is also not likely that 
metal enrichment abruptly ends the epoch of Pop~III star formation after the formation of the first stars.  As discussed in Section 2.2, 
it is possible for the photodissociating background radiation established by early generations of stars to slow the collapse of the primordial
gas, potentially delaying a large fraction of Pop~III star formation and metal enrichment until later times.  Also, as discussed in Section 3.2, the mixing of the first metals with the primordial
gas, especially within minihalos, may not occur efficiently.  Therefore, it is a distinct possibility that Pop~III star formation continues well after 
the formation of the first stars (e.g. Scannapieco et al. 2003; Tornatore et al. 2007; Trenti et al. 2009; Maio et al. 2010), and that substantial primordial star formation may be detectable in the first galaxies.  

\begin{figure}[b]
\sidecaption[t]
\includegraphics[scale=1.1]{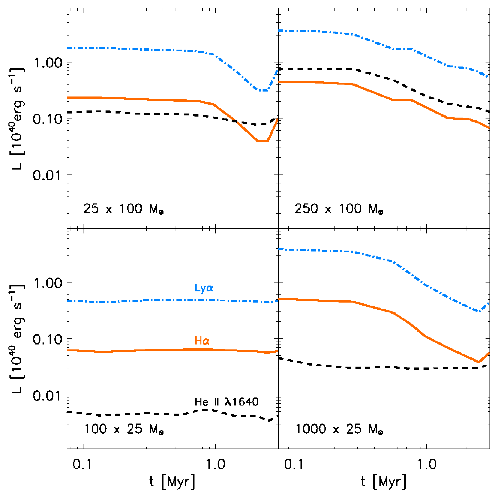}
%
%
\caption{The luminosity of Pop~III star clusters in a first galaxy at $z$ $\sim$ 12, as a function of the 
time from their formation, in three recombination lines: 
Ly$\alpha$ ({\it dot-dashed blue}), H$\alpha$ ({\it solid red}), and He~II 
$\lambda$1640 ({\it dashed black}). The four panels correspond to four different combinations of stellar IMF and total stellar mass; these are, clockwise 
from top-left: twenty-five 100 M$_{\odot}$ stars, two hundred fifty 100 M$_{\odot}$ stars, one thousand 25 M$_{\odot}$ stars, and one hundred 25 M$_{\odot}$ 
stars.  The relative luminosities of detected He~{\sc ii} and H~{\sc i} recombination lines can provide information about the stellar metallicity and IMF;
more massive and more metal-poor stars emit more high energy radiation that can ionize He~{\sc ii}, which leads to strong He~{\sc ii}~$\lambda$1640 emission relative to H$\alpha$ and Ly$\alpha$.  From Johnson et al. (2009).
}
\end{figure}

It is therefore critical to predict observable signatures of Pop~III star formation, in order that it can be identified in high redshift galaxies (e.g. Zackrisson et al. 2011).  
Some distinctive signatures derive from the high surface temperatures of primordial stars, which arise due to a relatively low opacity in the stellar interior.
This low opacity translates into a smaller radii $R_{\rm *}$ for primordial stars than for their metal-enriched counterparts.  
In turn, because stellar luminosity scales as $L_{\rm *}$ $\propto$ $R_{\rm *}^2$ $T_{\rm *}^4$, for a given luminosity the surface temperature $T_{\rm *}$ of
a primordial star will be higher than a metal-enriched star (e.g. Siess et al. 2002; Lawlor et al. 2008).  
For very massive primordial stars, the surface temperature is very high, roughly $\sim$ 10$^5$ K (Bromm et al. 2001; Schaerer 2002).
Owing to this high temperature, primordial stars emit copious high energy radiation, a relatively large fraction of which is able to ionize not only hydrogen (H~{\sc i}), but also helium (He~{\sc i} and He~{\sc ii}).  

As a substantial portion of the ionizing photons emitted from stars in early galaxies are absorbed by the relatively dense gas in the interstellar medium before escaping into the IGM 
(e.g. Wood \& Loeb 2000; Gnedin et al. 2008; Wise \& Cen 2009; Razoumov \& Sommer-Larsen 2010; Paardekooper et al. 2011; Yajima et al. 2011), the energy in these photons is reprocessed into emission lines arising from the recombination of the ionized species (e.g. Osterbrock \& Ferland 2006).\index{recombination}  For the case of primordial stars, because a relatively large fraction of the emitted radiation ionizes He~{\sc ii}, the photons emitted during the recombination of He~{\sc iii} to He~{\sc ii} produce
strong emission at characteristic wavelengths.  The most prominent recombination line emitted from such He~{\sc iii} regions, with a wavelength of 1640 ${\rm \AA}$, 
emerges from the radiative decay of the lone electron in this ion from the $n$ = 3 to the $n$ = 2 state\footnote{While photons are also emitted in transitions to the $n$ = 1 state, the IGM is optically thick to 
these photons before reionization due to absorption by neutral hydrogen, 
and so they are not expected to be observable from the first galaxies.}.  The most prominent emission lines from the recombination of ionized hydrogen in the H~{\sc ii} regions
surrounding primordial stars are the same as expected from metal-enriched stars, Ly$\alpha$ and H$\alpha$, which arise from the radiative decay from the $n$ = 2 $\to$ 1 and $n$ = 3 $\to$ 2 energy 
levels of hydrogen, respectively.  The key observational signature of primordial star formation, as opposed to metal-enriched star formation, is a relatively large ratio of the luminosity
emitted in the helium line, He~{\sc ii} $\lambda$1640, to that emitted in the hydrogen lines (see e.g. Tumlinson et al. 2001; Oh et al. 2001; Schaerer et al. 2003; Raiter et al. 2010).

\begin{figure}[t]
\sidecaption[t]
\includegraphics[scale=1.0]{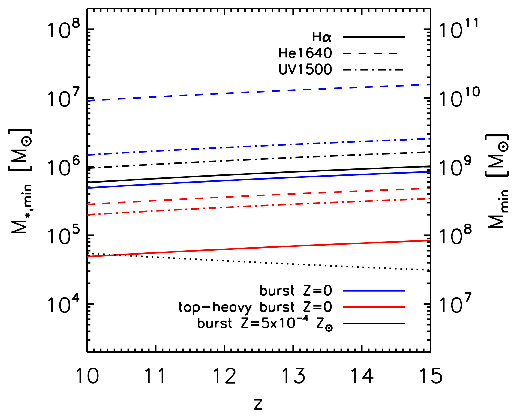}
%
%
\caption{Stellar masses $M_{\rm *,min}$ of the lowest mass starburst 
observable through the detection of the H$\alpha$ line ({\it solid curves}), 
the He~II $\lambda$1640 line ({\it dashed curves}), or the continuum at 1500~\AA ({\it dash- 
dotted curves}) with JWST, assuming an exposure time of 10$^6$ s 
and a signal-to-noise ratio of S/N = 10. Stellar masses derived 
from the Schaerer (2003) zero-metallicity starbursts with a standard Salpeter-like
IMF, zero-metallicity starbursts with a top-heavy IMF, and
low-metallicity starbursts, are shown in blue, red, and 
black, respectively. The right axis shows the masses $M_{\rm min}$ = 10$^3$ M$_{\rm *,min}$ of 
halos expected to host a starburst with stellar mass $M_{\rm *,min}$.  The dotted curve 
shows the mass of a first galaxy-sized halo, with virial temperature $T_{\rm vir}$ = 10$^4$ K.  From Pawlik et al. (2011).}
\end{figure}

Figure 18 shows the luminosity emitted in each of the three recombination lines mentioned above from an instantaneous burst of Pop~III star formation in a first galaxy formed in a halo of mass $M_{\rm h}$ $\sim$ 
10$^8$ M$_{\odot}$ at $z$ $\sim$ 12, as gleaned from cosmological radiative transfer simulations (Johnson et al. 2009).  Each of the panels shows the line luminosities for a different combination of the characteristic 
stellar mass of the stars (either 25 or 100 M$_{\odot}$) and of the total stellar mass (either 2,500 or 25,000 M$_{\odot}$).  Even assuming such large characteristic masses for Pop~III stars
and that such a large fraction (either $\sim$ 1 or $\sim$ 10 percent) of the gas in the first galaxies is converted into stars, the luminosities of the recombination lines are likely to be too dim to detect with telescopes
in the near future.  To see this, we can estimate the total flux $F$ that would be in these lines at $z$ = 0 as 

\begin{eqnarray}
F &  = & \frac{L }{4\pi D_{\rm L}^2(z)} \nonumber \\ 
                &  \simeq & 10^{-20}  \left(\frac{L}{10^{40} {\rm erg\, s^{-1}}}\right)
                \left(\frac{1+z}{10}\right)^{-2} {\rm erg \: s}^{-1} \: {\rm cm}^{-2} \mbox{\ ,} 
\end{eqnarray}
where $L$ is the luminosity in a given line and $D_{\rm L}$ is the luminosity distance to redshift $z$.  At $z$ $\ge$ 10, even the most luminous line, Ly$\alpha$, would be seen at $z$ = 0 with a flux of 
of $\le$ 4 $\times$ 10$^{-20}$ erg s$^{-1}$ cm$^{-2}$, which is well below the flux limit of $\sim$ 2 $\times$ 10$^{-19}$ erg s$^{-1}$ cm$^{-2}$ of surveys planned for the 
JWST (Gardner et al. 2006; Windhorst et al. 2006).  \index{James Webb Space Telescope}

Instead of the first galaxies, hosted in halos with masses of $\sim$ 10$^8$ M$_{\odot}$ at $z$ $\ge$ 10, it thus appears likely that observations in the next 
decade may reveal somewhat more developed galaxies hosted in more massive halos (e.g. Barkana \& Loeb 2000; Ricotti et al. 2008; Johnson et al. 2009; Pawlik et al. 2011), although there 
is the possibility of detecting less developed galaxies if their flux is magnified by gravitational lensing (see e.g. Zackrisson 2011).    
As shown in Figure 19, the JWST is predicted to be capable of detecting both He~{\sc ii}~$\lambda$1640 and H$\alpha$ from metal-free starbursts in halos with masses $\ge$ 3$\times$ 10$^{8}$ M$_{\odot}$,
if the IMF is very top-heavy.  Though, if the typical mass of Pop~III stars is $<$ 50 M$_{\odot}$, it is likely that He~{\sc ii}~$\lambda$1640 will only be detectable from significantly more massive stellar clusters, 
expected to form in similarly more massive
halos.  However, because more massive halos are formed from the mergers of smaller halos which themselves may have hosted star formation, it may be predominantly metal-enriched Pop~II stars that form in the 
galaxies which will be detected by the JWST (e.g. Johnson et al. 2008).\footnote{It is also likely that other, complementary next generation facilities, such as the {\it Atacama Large Millimeter Array} (e.g. Combes 2010),
will detect only metal-enriched star-forming galaxies.}

That said, there is the possibility that substantial Pop~III star formation takes place even well after the epoch of the first galaxies (i.e. at $z$ $<$ 10), either due to  
inefficient mixing of primordial and metal-enriched gas (e.g. Jimenez \& Haiman 2006; Pan \& Scalo 2007; Wyithe \& Cen 2007; Dijkstra \& Wyithe 2007; Cen 2010) or to the collapse
of primordial gas into late-forming atomic cooling halos (e.g. Tornatore et al. 2007; Trenti et al. 2009; Johnson 2010).  Pop~III star formation at such late times could be detected 
more easily, in large part because the emission line flux increases strongly with decreasing redshift, as shown in equation (44).  \index{Population III}
However, at such low redshifts the background ionizing radiation field that builds up during reionization can strongly inhibit the infall of primordial gas into halos, limiting the
amount of Pop~III star formation that can occur even in metal-free galaxies (e.g. Efstathiou 1992; Gnedin 2000; Tassis et al. 2003; Dijkstra et al. 2004).

\section{Summary and Conclusion}
In this Chapter we have discussed a wide range of the physical processes that must be accounted for in the theoretical modeling of the first galaxies.
We have made important distinctions between the formation of the first stars in minihalos and star formation in the atomic cooling halos hosting the first galaxies,
highlighting how the cooling properties of the gas assembled into the first galaxies are altered by high energy radiation and by the injection of heavy elements
from the first supernovae.  While this has not been a complete review of the theory of the formation of the first galaxies, it has hopefully served to illustrate, 
from basic principles where possible, much of the physics that comes into play in their study.  The reader is referred to the 
many excellent articles in the bibliography below for more in-depth study on the topic.

In closing, it is critical to point out that without accounting for all of the effects we have discussed together, one is left with an incomplete understanding of the first galaxies.
For instance, as we have seen, the radiation from the first stars can ionize the gas and trigger HD cooling, but it can also easily destroy H$_{\rm 2}$ and HD molecules. 
As well, while the first supernovae may enrich much of the gas from which the first   
galaxies form to a level above the critical metallicity needed for low-mass Pop~II star formation, much of the dense gas in minihalos may not be efficiently mixed with the metal-enriched ejecta.
Similarly, black holes may only form by direct collapse in rare regions in which the LW background radiation field is elevated, but the same stars which likely produce this 
radiation may also enrich the gas when they explode as supernovae, possibly precluding this mode of black hole formation.
A complete and consistent picture of the formation of the first galaxies only emerges when accounting for star and black hole formation, metal enrichment, 
and radiative feedback all together in the full cosmological context.

Making this task especially daunting is the range of scales that must be taken into account.  
The gas clouds which collapse to form stars are on sub-parsec scales, metal-enrichment from the first supernovae 
occurs on parsec to kiloparsec scales, and the radiation emitted by the first stars can impact regions on kiloparsec or even megaparsec scales.  
Thus, simulations must ultimately resolve an enormous range of scales in order to capture all of the important physical processes that come into play.
While we have introduced the results of numerous analytical calculations and simulations, none of them alone captures all of the processes we have discussed
simultaneously.  Indeed, this stands as one of the primary challenges to making detailed predictions of the nature of the first galaxies.  

\begin{acknowledgement}
The author is grateful to the editors for the invitation to contribute this Chapter, as well as to Bhaskar Agarwal, Volker Bromm, Umberto Maio, and Eyal Neistein for helpful comments on an earlier draft of this work. Credit also goes to Chalence Safranek-Shrader for identifying an error (now corrected) in equation (29), 
as well as to Daisuke Nakauchi for identifying an inconsistency (now corrected) between equations (25) and (26).
\end{acknowledgement}


\begin{thebibliography}{199}%

\bibitem{a1}Abel, T., Anninos, P., Zhang, Y., Norman, M.~L. 1997, NewAR, 2, 181--207
\bibitem{a1}Abel, T., Bryan, G.~L., Norman, M.~L. 2002, Sci, 295, 93--98
\bibitem{a1}Abel, T., Wise, J.~H., Bryan, G.~L. 2007, ApJ, 659, L87--L90
\bibitem{a1}Ahn, K., Shapiro, P.~R., Iliev, I.~T., Mellema, G., Pen, U.-L. 2009, ApJ, 695, 1430--1445
\bibitem{a1}Alvarez, M.~A., Bromm, V., Shapiro, P.~R. 2006, ApJ, 639, 621--632
\bibitem{a1}Alvarez, M.~A., Wise, J.~H., Abel, T. 2009, ApJ, 701, L133--L137
\bibitem{a1}Aykutalp, A., Spaans, M. 2011, ApJ, in press (arXiv:1105.5158)
\bibitem{a1}Barkana, R., Loeb, A. 2000, ApJ, 531, 613--623
\bibitem{a1}Barkana, R., Loeb, A. 2001, Phys. Rep., 349, 125--238  
\bibitem{a1}Begelman, M.~C., Volonteri, M., Rees, M.~J. 2006, MNRAS, 370, 289--298
\bibitem{a1}Begelman, M.~C., Shlosman, I. 2009, ApJ, 702, L5--L8
\bibitem{a1}Bondi, H. 1952, MNRAS, 112, 195--204
\bibitem{a1}Bromm, V., Coppi, P.~S., Larson, R.~B. 2002, ApJ, 564, 23--51 
\bibitem{a1}Bromm, V., Ferrara, A., Coppi, P.~S., Larson, R.~B. 2001, MNRAS, 328, 969--976
\bibitem{a1}Bromm, V., Kudritzki, R.~P., Loeb, A. 2001, ApJ, 552, 464--472 
\bibitem{a1}Bromm, V., Larson, R.~B. 2004, ARA\&A, 42, 79--118
\bibitem{a1}Bromm, V., Loeb, A. 2003a, ApJ, 596, 34--46
\bibitem{a1}Bromm, V., Loeb, A. 2003b, Nat, 425, 812--814
\bibitem{a1}Bromm, V., Yoshida, N. 2011, ARA\&A, 49, 373--407
\bibitem{a1}Bromm, V., Yoshida, N., Hernquist, L. 2003, ApJ, 596, L135-L138
\bibitem{a1}Bromm, V., Yoshida, N., Hernquist, L., McKee, C.~F. 2009, Nat, 459, 49--54
\bibitem{a1}Burbidge, E., Burbidge, G.~R., Fowler, W.~A., Hoyle, F. 1957, Rev. Mod. Phys., 1957, 547--650
\bibitem{a1}Cen, R. 2010, ApJ, 725, 115--120
\bibitem{a1}Cen, R., Riquelme, M.~A. 2008, ApJ, 674, 644-652
\bibitem{a1}Cherchneff, I., Dwek, E. 2010, ApJ, 713, 1--24 
\bibitem{a1}Ciardi, B., Ferrara, A. 2005, Space Sci. Rev., 116, 625--705
\bibitem{a1}Ciardi, B., Ferrara, A., Abel, T. 2000, ApJ, 533, 594--600
\bibitem{a1}Clark, P.~C., Glover, S.~C.~O., Smith, R.~J., Greif, T.~H., Klessen, R.~S., Bromm, V. 2011a, Sci, 331, 1040--1042 
\bibitem{a1}Clark, P.~C., Glover, S.~C.~O., Klessen, R.~S. 2008, ApJ, 672, 757--764
\bibitem{a1}Clark, P.~C., Glover, S.~C.~O., Klessen, R.~S., Bromm, V. 2011b, ApJ, 727, 110-127 
\bibitem{a1}Combes, F. 2010, Am. Inst. Phys. Conf. Proc., 1294, 9--16
\bibitem{a1}de Avillez, M.~A., Mac Low, M.-M. 2002, ApJ, 581, 1047--1060 
\bibitem{a1}de Souza, R.~S., Rodrigues, L.~F.~S., Opher, R. 2011, MNRAS, 410, 2149--2155 
\bibitem{a1}Dijkstra, M., Haiman, Z., Mesinger, A., Wyithe, J.~S.~B. 2008, MNRAS, 391, 1961--1972 
\bibitem{a1}Dijkstra, M., Haiman, Z., Rees, M.~J., Weinberg, D.~H. 2004, ApJ, 601, 666--675 
\bibitem{a1}Dijkstra, M., Wyithe, J.~S.~B. 2007, MNRAS, 379, 1589--1598
\bibitem{a1}Efstathiou, G. 1992, MNRAS, 256, 43--47
\bibitem{a1}Ferrara, A., Pettini, M., Shchekinov, Y. 2000, MNRAS, 319, 539--548 
\bibitem{a1}Flower D.R., Le Bourlot J., Pineau des For{\^{e}}ts G., Roueff E. 2000, MNRAS, 314, 753--758 
\bibitem{a1}Frebel, A., Johnson, J.~L., Bromm, V. 2007, MNRAS, 380, L40--L44 
\bibitem{a1}Frebel, A., Johnson, J.~L., Bromm, V. 2009, MNRAS, 392, L50--L54 
\bibitem{a1}Fryer, C.~L., Woosley, S.~E., Heger, A. 2001, ApJ, 550, 372--382
\bibitem{a1}Galli, D., Palla, F. 1998, A\&A, 335, 403--420 
\bibitem{a1}Gao, L., Theuns, T., Frenk, C.~S., Jenkins, A., Helly, J.~C., Navarro, J., Springel, V., White, S.~D.~M. 2010, MNRAS, 403, 1238--1295 
\bibitem{a1}Gardner, J.~P., et al. 2006, Space Sci. Rev., 123, 485--606 
\bibitem{a1}Glover, S.~C.~O. 2005, Space Sci. Rev., 117, 445--508
\bibitem{a1}Glover, S.~C.~O., Brand, P.~W.~J.~L. 2001, MNRAS, 321, 385--397
\bibitem{a1}Gnedin, N.~Y. 2000, ApJ, 542, 535--541 
\bibitem{a1}Gnedin, N.~Y., Kravtsov, A.~V., Chen, H.-W. 2008, ApJ, 672, 765--775 
\bibitem{a1}Greif, T.~H., Bromm, V. 2006, MNRAS, 373, 128--138 
\bibitem{a1}Greif, T.~H., Glover, S.~C.~O., Bromm, V., Klessen, R.~S. 2010, ApJ, 716, 510--520
\bibitem{a1}Greif, T.~H., Johnson, J.~L., Bromm, V., Klessen, R.~S. 2007, ApJ, 670, 1--14 
\bibitem{a1}Greif, T.~H., Johnson, J.~L., Klessen, R.~S., Bromm, V. 2008, MNRAS, 387, 1021--1036
\bibitem{a1}Greif, T.~H., Springel, V., White, S.~D.~M., Glover, S.~C.~O., Clark, P.~C., Smith, R.~J., Klessen, R.~S., Bromm, V. 2011, ApJ, submitted (arXiv:1101.5491)
\bibitem{science-contrib}Haiman, Z. 2009, Astrophysics in the Next Decade, 385-418 (arXiv:0809.3926)
\bibitem{a1}Haiman, Z., Rees, M.~J., Loeb, A. 1997, ApJ, 476, 458--463 
\bibitem{a1}Heger, A., Fryer, C.~L., Woosley, S.~E., Langer, N., Hartmann, D.~H. 2003, ApJ, 591, 288--300
\bibitem{a1}Heger, A., Woosley, S.~E. 2002, ApJ, 567, 532--543 
\bibitem{a1}Heger, A., Woosley, S.~E. 2010, ApJ, 724, 341--373
\bibitem{a1}Jappsen, A.-K., Klessen, R.~S., Glover, S.~C.~O., Mac Low, M.-M. 2009a, ApJ, 694, 1065--1074
\bibitem{a1}Jappsen, A.-K., Mac Low, M.-M., Glover, S.~C.~O., Klessen, R.~S., Kitsionas, S. 2009b,  ApJ, 694, 1161--1170 
\bibitem{a1}Jasche, J., Ciardi, B., En{\ss}lin, T.~A. 2007, MNRAS, 380, 417--429
\bibitem{a1}Jimenez, R., Haiman, Z. 2006, Nat, 440, 501--504 
\bibitem{a1}Johnson, J.~L. 2010, MNRAS, 404, 1425--1436
\bibitem{a1}Johnson, J.~L., Bromm, V. 2006, MNRAS, 366, 247--256
\bibitem{a1}Johnson, J.~L., Bromm, V. 2007, MNRAS, 374,  1557--1568
\bibitem{a1}Johnson, J.~L., Greif, T.~H., Bromm, V. 2008, MNRAS, 388, 26--38
\bibitem{a1}Johnson, J.~L., Greif, T.~H., Bromm, V., Klessen, R.~S., Ippolito, J. 2009, MNRAS, 399, 37--47
\bibitem{a1}Johnson, J.~L., Khochfar, S. 2011, MNRAS, 413, 1184--1191 
\bibitem{a1}Johnson, J.~L., Khochfar, S., Greif, T.~H., Durier, F. 2011, MNRAS, 410, 919-933 
\bibitem{a1}Karlsson, T. 2005, A\&A, 439, 93--106 
\bibitem{a1}Karlsson, T., Bromm, V., Bland-Hawthorn, J. 2011, Rev. Mod. Phys., submitted (arXiv:1101.4024) 
\bibitem{a1}Karlsson, T., Johnson, J.~L., Bromm, V. 2008, ApJ, 679, 6--16 
\bibitem{a1}Kitayama, T., Ikeuchi, S. 2000, ApJ, 529, 615--634
\bibitem{a1}Kitayama, T., Susa, H., Umemura, M., Ikeuchi, S. 2001, MNRAS, 326, 1353--1366 
\bibitem{a1}Kitayama, T., Yoshida, N. 2005, ApJ, 630, 675--688 
\bibitem{a1}Klessen, R.~S., Lin, D.~N. 2003, Phys. Rev. E, 67
\bibitem{a1}Komiya, Y., Habe, A., Suda, T., Fujimoto, M.~Y. 2010, ApJ, 717, 542--561
\bibitem{a1}Koushiappas, S.~M., Bullock, J.~S., Dekel, A. 2004, MNRAS, 354, 292--304
\bibitem{a1}Kuhlen, M., Madau, P. 2005, MNRAS, 363, 1069--1082 
\bibitem{a1}Kulsrud, R.~M., Cen, R., Ostriker, J.~P., Ryu, D. 1997, ApJ, 480, 481--491 
\bibitem{a1}Larson, R.~B. 2005, MNRAS, 359, 211--222 
\bibitem{a1}Lawlor, T.~M., Young, T.~R., Johnson, T.~A., MacDonald, J. 2008, MNRAS, 384, 1533--1543 
\bibitem{a1}Leitherer, C., Schaerer, D., Goldader, J.~D., Gonz{\' a}lez Delgado, R.~M., Robert, Carmelle, Kune, D.~F., de Mello, D.~F., Devost, D., Heckman, T.~M. 1999, ApJ, Suppl. 123, 3--40
\bibitem{a1}Lodato, G., Natarajan, P. 2006, MNRAS, 371, 1813--1823
\bibitem{a1}Mac Low, M.-M., Klessen, R.~S. 2004, Rev. Mod. Phys., 76, 125--194 
\bibitem{a1}Machacek, M.~E., Bryan, G.~L., Abel, T. 2001, ApJ, 548, 509--521 
\bibitem{a1}Machida, M.~N., Tomisaka, K., Nakamura, F., Fujimoto, M.~Y. 2005, ApJ, 622, 39--57 
\bibitem{a1}Mackey, J., Bromm, V., Hernquist, L. 2003, ApJ, 586, 1--11 
\bibitem{a1}Madau, P., Ferrara, A., Rees, M.~J. 2001, ApJ, 555, 92--105
\bibitem{a1}Madau, P., Kuhlen, M., Diemand, J., Moore, B., Zemp, M., Potter, D., Stadel, J. 2008, ApJ, 689, L41--L44 
\bibitem{a1}Maio, U., Dolag, K., Ciardi, B., Tornatore, L. 2007, MNRAS, 379, 963--973
\bibitem{a1}Maio, U., Ciardi, B., Dolag, K., Tornatore, L., Khochfar, S. 2010, MNRAS, 407, 1003--1015 
\bibitem{a1}Maio, U., Khochfar, S., Johnson, J.~L., Ciardi, B. 2011, MNRAS, in press (arXiv:1011.3999)
\bibitem{a1}Mayer, L., Kazantzidis, S., Escala, A., Callegari, S. 2010, Nat, 466, 1082--1084 
\bibitem{a1}McGreer, I.~D., Bryan, G.~L. 2008, ApJ, 685, 8--20 
\bibitem{a1}McKee, C.~F., Tan, J.~C. 2008, ApJ, 681, 771--797 
\bibitem{a1}Mesinger, A., Bryan, G.~L., Haiman, Z. 2006, ApJ, 648, 835--851
\bibitem{a1}Milosavljevi{\' c}, M., Bromm, V., Couch, S.~M., Oh, S.~P. 2009, ApJ, 698, 766--780 
\bibitem{a1}Mori, M., Ferrara, A., Madau, P. 2002, ApJ, 571, 40--55
\bibitem{a1}Murray, S.~D., White, S.~D.~M., Blondin, J.~M., Lin, D.~N.~C. 1993, ApJ, 407, 588--596
\bibitem{a1}Nagakura, T., Omukai, K. 2005, MNRAS, 364, 1378--1386  
\bibitem{a1}Nakamura, F., Umemura, M. 2002, ApJ, 569, 549--557
\bibitem{a1}Nozawa, T., Kozasa, T., Umeda, H., Maeda, K., Nomoto, K. 2003, ApJ, 598, 785--803 
\bibitem{a1}Oh, S.~P. 2001, ApJ, 553, 499--512 
\bibitem{a1}Oh, S.~P., Haiman, Z. 2002, ApJ, 569, 558--572
\bibitem{a1}Oh, S.~P., Haiman, Z., Rees, M.~J. 2001, ApJ, 553, 73--77 
\bibitem{a1}Omukai, K., Nishi, R. 1999, ApJ, 518, 64--68 
\bibitem{a1}Omukai, K., Palla, F. 2003, ApJ, 589, 677--687 
\bibitem{a1}Omukai, K., Schneider, R., Haiman, Z. 2008, ApJ, 686, 801--814 
\bibitem{a1}Omukai, K., Tsuribe, T., Schneider, R., Ferrara, A. 2005, ApJ, 626, 627--643 
\bibitem{a1}O'Shea, B.~W., Norman, M.~L. 2008, ApJ, 673, 14--33 
\bibitem{science-mono}Osterbrock, D.~O., Ferland, G.~J. 2006, Astrophysics of Gaseous Nebulae and Active Galactic Nuclei. University Science Books, Sausalito  
\bibitem{a1}Paardekooper, J.-P., Pelupessy, F.~I., Altay, G., Kruip, C. 2011,  A\&A in press (arXiv:1104.3584)
\bibitem{a1}Padoan, P., Nordlund, \AA., Kritsuk, A.~G., Norman, M.~L., Pak Shing, L. 2007, ApJ, 661, 972--981
\bibitem{a1}Pan, L., Scalo, J. 2007, ApJ, 654, L29--L32
\bibitem{a1}Park, K., Ricotti, M. 2010, ApJ, submitted (arXiv:1006.1302)
\bibitem{a1}Pawlik, A.~H., Milosavljevi{\' c}, M., Bromm, V. 2011, ApJ, 731, 54--70
\bibitem{a1}Pelupessy, F.~I., Di Matteo, T., Ciardi, B. 2007, ApJ, 665, 107--119
\bibitem{a1}Prieto, J., Padoan, P., Jimenez, R., Infante, L. 2011, ApJ, 731, L38--L43 
\bibitem{a1}Pringle, J.~E. 1981, ARA\&A, 19, 137--162 
\bibitem{a1}Raiter, A., Schaerer, D., Fosbury, R.~A.~E. 2010, A\&A 523, 64 
\bibitem{a1}Razoumov, A.~O., Sommer-Larsen, J. 2010, ApJ, 710, 1239--1246
\bibitem{a1}Read, J.~I., Pontzen, A.~P., Viel, M. 2006, MNRAS, 371, 885--897 
\bibitem{a1}Regan, J.~A., Haehnelt, M.~G. 2009, MNRAS, 396, 343--353
\bibitem{a1}Ricotti, M., Gnedin, N.~Y., Shull, J.~M. 2001, ApJ, 560, 580--591
\bibitem{a1}Ricotti, M., Gnedin, N.~Y., Shull, J.~M. 2008, ApJ, 685, 21--39 
\bibitem{a1}Ricotti, M., Ostriker, J.~P. 2004, MNRAS, 375, 547--562 
\bibitem{a1}Ricotti, M., Shull, J.~M. 2000, ApJ, 542, 548--558
\bibitem{a1}Ripamonti, E. 2007,  MNRAS, 376, 709--718 
\bibitem{a1}Rodrigues, L.~F.~S., de Souza, R.~S., Opher, R. 2010, MNRAS, 406, 482--485
\bibitem{a1}Safranek-Shrader, C., Bromm, V., Milosavljeci{\' c}, M. 2010, ApJ, 723, 1568--1582 
\bibitem{a1}Santoro, F., Shull, J.~M. 2006, ApJ, 643, 26--37 
\bibitem{a1}Scannapieco, E., Schneider, R., Ferrara, A. 2003, ApJ, 589, 35--52 
\bibitem{a1}Schaerer, D. 2002, A\&A, 382, 28--42
\bibitem{a1}Schaerer, D. 2003, A\&A, 397, 527--538
\bibitem{a1}Schleicher, D.~R.~G., Banerjee, R., Sur, S., Arshakian, T.~G., Klessen, R.~S., Beck, R., Spaans, M. 2010a, A\&A, 522, A115--A124 
\bibitem{a1}Schleicher, D.~R.~G., Galli, D., Glover, S.~C.~O., Banerjee, R., Palla, F., Schneider, R., Klessen, R.~S. 2009, ApJ, 703, 1096--1106 
\bibitem{a1}Schleicher, D.~R.~G., Spaans, M., Glover, S.~C.~O. 2010b, ApJ, 712, L69--L72 
\bibitem{a1}Schneider, R., Ferrara, A., Salvaterra, R. 2004, MNRAS, 351, 1379--1386
\bibitem{a1}Schneider, R., Omukai, K. 2010, MNRAS, 402, 429--435
\bibitem{a1}Schneider, R., Omukai, K., Inoue, A.~K., Ferrara, A. 2006, MNRAS, 369, 1437--1444 
\bibitem{a1}Shang, C., Bryan, G.~L., Haiman, Z. 2010, MNRAS, 402, 1249--1262 
\bibitem{a1}Shapiro, P.~R., Kang, H. 1987, ApJ, 318, 32--65
\bibitem{a1}Siess, L., Livio, M., Lattanzio, J. 2002, ApJ, 570, 329--343 
\bibitem{a1}Silk, J., Langer, M. 2006, MNRAS, 371, 444--450 
\bibitem{a1}Smith, B.~D., Sigurdsson, S. 2007,  ApJ, 661, L5-L8
\bibitem{a1}Smith, B.~D., Turk, M.~J., Sigurdsson, S., O'Shea, B.~W., Norman, M.~L. 2009, ApJ, 691, 441--451
\bibitem{a1}Spaans, M., Silk, J. 2006, ApJ, 652, 902--906
\bibitem{a1}Stacy, A., Bromm, V. 2007, MNRAS, 382, 229--238 
\bibitem{a1}Stacy, A., Greif, T.~H., Bromm, V. 2010, MNRAS, 403, 45--60
\bibitem{science-mono}Stahler, S.~W., Palla, F. 2004, The Formation of Stars, Wiley 
\bibitem{a1}Stahler, S.~W., Shu, F.~H., Taam, R.~E. 1980, ApJ, 241, 637--654
\bibitem{a1}Stecher, T.~P., Williams, D.~A. 1967, ApJ, 149, L29--L30
\bibitem{a1}Suda, T., Aikawa, M., Machida, M.~N., Fujimoto, M.~Y., Iben, I. 2004, ApJ, 611, 476--493 
\bibitem{a1}Tan, J.~C., McKee, C.~F. 2004, ApJ, 603, 383--400
\bibitem{a1}Tassis, K., Abel, T., Bryan, G.~L., Norman, M.~L. 2003, ApJ, 587, 13--24
\bibitem{a1}Tenorio-Tagle, G. 1996, ApJ, 111, 1641--1650
\bibitem{a1}Tominaga, N., Umeda, H., Nomoto, K. 2007, ApJ, 660, 516--540
\bibitem{a1}Tornatore, L., Ferrara, A., Schneider, R. 2007,  MNRAS, 382, 945--950 
\bibitem{a1}Trenti, M., Stiavelli, M. 2009, ApJ, 694, 879--892
\bibitem{a1}Trenti, M., Stiavelli, M., Shull, J.~M. 2009, ApJ, 700, 1672--1679
\bibitem{a1}Tumlinson, J. 2006, ApJ, 641, 1--20 
\bibitem{a1}Tumlinson, J., Giroux, M.~L., Shull, J.~M. 2001, ApJ, 550, L1--L5
\bibitem{a1}Turk, M.~J., Abel, T., O'Shea, B. 2009, Sci, 325, 601--605
\bibitem{a1}Uehara, H., Inutsuka, S. 2000, ApJ, 531, L91--L94
\bibitem{a1}Vasiliev, E.~O., Shchekinov, Y.~A. 2006, Astron. Rep., 50, 778--784
\bibitem{a1}Vasiliev, E.~O., Vorobyov, E.~I. 2008, A\&A, 489, 505--515
\bibitem{a1}Wada, K., Venkatesan, A. 2003, ApJ, 591, 38--42
\bibitem{a1}Whalen, D., Hueckstadt, R.~M., McConkie, T.~O. 2010, ApJ, 712, 101--111
\bibitem{a1}Whalen, D., van Veelen, B., O'Shea, B.~W., Norman, M.~L. 2008, ApJ, 682, 49--67 
\bibitem{a1}Windhorst, R.~A., Cohen, S.~H., Jansen, R.~A., Conselice, C., Yan, H. 2006, NewAR, 50, 113--120
\bibitem{a1}Wise, J.~H., Abel, T. 2007a, ApJ, 671, 1559--1567 
\bibitem{a1}Wise, J.~H., Abel, T. 2007b, ApJ, 665, 899--910 
\bibitem{a1}Wise, J.~H., Abel, T. 2008, ApJ, 685, 40--56
\bibitem{a1}Wise, J.~H., Cen, R. 2009, ApJ, 693, 984--999 
\bibitem{a1}Wise, J.~H., Turk, M.~J., Norman, M.~L., Abel, T. 2011, ApJ, submitted (arXiv:1011.2632)
\bibitem{a1}Wolcott-Green, J., Haiman, Z. 2011, MNRAS, 412, 2603--2616 (2011)
\bibitem{a1}Wood, K., Loeb, A. 2000, ApJ, 545, 86--99
\bibitem{a1}Wyithe, J.~S.~B., Cen, R. 2007, ApJ, 659, 890--907
\bibitem{a1}Wyithe, J.~S.~B., Loeb, A. 2003, ApJ, 586, 693--708
\bibitem{a1}Xu, H., O'Shea, B.~W., Collins, D.~C., Norman, M.~L., Li, H., Li, S. 2008, ApJ, 688, L57--L60
\bibitem{a1}Yajima, H., Choi, J.-H., Nagamine, K. 2011, MNRAS, 412, 411--422
\bibitem{a1}Yoshida, N. 2006, NewAR, 50, 19--23
\bibitem{a1}Yoshida, N., Abel, T., Hernquist, L., Sugiyama, N. 2003, ApJ, 592, 645--663
\bibitem{a1}Yoshida, N., Oh, S.~P., Kitayama, T., Hernquist, L. 2007a, ApJ, 663, 687--707
\bibitem{a1}Yoshida, N., Omukai, K., Hernquist, L. 2007b, ApJ, 667, L117--L120
\bibitem{a1}Yoshida, N., Omukai, K., Hernquist, L. 2008, Sci, 321, 669--671
\bibitem{a1}Zackrisson, E. 2011, Conf. Proc. (arXiv:1101.4033)
\bibitem{a1}Zackrisson, E., Rydberg, C.-E., Schaerer, D., Ostlin, G., Tuli, M. 2011, ApJ, submitted (arXiv:1105.0921) 




\end{thebibliography}
\end{document}